\documentclass[11pt]{article}
\usepackage{epsfig}
\newcommand{\mbf}[1]{\mbox{\boldmath $#1$}}
\usepackage{graphicx}% Include figure files
\usepackage{amsmath}
\usepackage{dsfont}
\newcommand{\bk}{\mbf{k}}
\newcommand{\bq}{\mbf{q}}
\newcommand{\bp}{\mbf{p}}
\newcommand{\bb}{\mbf{b}}
\newcommand{\disc}{\textrm{disc}}
\newcommand{\cA}{{\cal A}}
\newcommand{\cT}{{\cal T}}
\newcommand{\cF}{{\cal F}}
\newcommand{\cN}{{\cal N}}
\newcommand{\cM}{{\cal M}}
\newcommand{\cD}{{\cal D}}
\newcommand{\cK}{{\cal K}}
\newcommand{\cG}{{\cal G}}
\newcommand{\tF}{\tilde{\cal F}}
\newcommand{\tA}{\tilde{\cal A}}
\newcommand{\tT}{\tilde{\cal T}}

\title{\bf AGK Cutting Rules and Multiple Scattering\\in Hadronic Collisions} 
\author{J.Bartels$^{a}$, M.Salvadore$^{a,b}$, and G.P.Vacca$^{b}$ \\
$^a$ II Inst. f. Theor. Physik, Univ. of Hamburg, Germany and\\
$^b$Dipartimento di Fisica and INFN, Univ. of Bologna, Italy }
\date{}
\def\beq{\begin{equation}}
\def\eeq{\end{equation}}
\def\bea{\begin{eqnarray}}
\def\eea{\end{eqnarray}}

\def\o{\omega}
\begin{document}
\maketitle
\medskip
\noindent
{\bf Abstract:}\\
We discuss the AGK rules for the exchange of an arbitrary number 
of reggeized gluons in perturbative QCD in the high energy limit.
Results include the cancellation  of corrections to single jet and double
jet inclusive cross sections,  both for hard and soft rescattering 
contributions.

\section{Introduction}
More than 30 years ago, Abramovsky, Gribov, and Kancheli in their pioneering 
paper ~\cite{AGK} have pointed out that, for high energy hadron hadron 
scattering, the multiple exchange of Pomerons with intercept at or close to 
unity leads to observable effects in multiparticle final states. 
As an example, double Pomeron exchange predicts fluctuations of 
the rapidity densities of produced particles; furthermore, in double 
inclusive particle production multi-Pomeron exchange leads to long range 
rapidity correlations. All these results suggests that, at very high energies,
hadron hadron scattering might exhibit some sort of critical behavior.  
As to the theoretical basis of these predictions, only fairly general 
properties of Regge theory have been used, especially features 
of particle-reggeon vertices which are quite independent of any special 
underlying quantum field theory. In particular, no reference has been made to 
QCD.

In recent years, the investigation of the small-$x$ region of hard 
scattering processes has lead to interest in the BFKL
Pomeron~\cite{BFKL}
which, at short distances, describes the exchange of vacuum quantum
numbers at high energies. In connection with saturation and the color
glass condensate, both in DIS at HERA and in heavy ion collisions at
RHIC also the 
multiple exchange of BFKL Pomerons is being addressed; a convenient framework 
for this is the nonlinear Balitsky-Kovchegov (BK) equation. 
At hadron colliders, there is more and more evidence 
~\cite{SS} that at high energies 
multiple interactions of partons cannot be neglected.   
All this naturally suggests to perform the AGK 
analysis in the framework of perturbative QCD where the Pomeron 
is described by the sum of BFKL ladder diagrams. In particular, 
it will be of interest to understand how the presence of multiple 
scattering affects, for example, the multiplicity of jets in the final state 
or the rapidity correlation between two jets.
A special motivation for investigating multiple scattering comes from the 
interest in saturation: most direct signals of the existence of saturation 
are expected to be seen in specific features of final states. 
Use of the AGK rules in pQCD has already been made in several papers
\cite{Braun,Kovchegov}.

The theoretical basis for studying these rules in perturbative QCD 
has been laid down in ~\cite{BR}. A clean theoretical environment for 
studying high energy scattering processes in pQCD is provided by virtual 
photons where the mass of the photon defines the momentum scale of the QCD 
coupling.
The coupling of two BFKL Pomerons to a virtual photon has been analysed in
~\cite{BW}; a generalization to three BFKL ladders has been investigated in 
~\cite{BE}. It is the structure of these couplings of BFKL Pomerons to 
the external photon which provides the basis for performing the AGK analysis
in pQCD. The most interesting applications of the pQCD AGK-analysis 
include deep inelastic electron proton scattering or multijet production in 
$pp$ collisions: in both cases one needs multi-gluon couplings to the 
proton. In the absence of any first-principle calculation, these couplings 
have to be modelled, similarly to the initial conditions of the parton 
densities. The perturbative analysis of the coupling of two (and more) BFKL 
Pomerons to virtual photons suggests that these couplings are not 
arbitrary: they obey certain symmetry requirements, which can 
be expected to be valid also beyond perturbation theory. We therefore believe 
that our understanding of the pQCD couplings will help to model, for example 
in proton proton scattering, the coupling of $n$ gluons to the proton.

In this paper we therefore begin with a brief review of the main results 
of ~\cite{BR}, and we point out which features of
the couplings are crucial 
for obtaining the AGK results\footnote{The issue of AGK rules
in connection with multiparticle interaction has already been addressed
in a different way in \cite{RT} without investigation of the
cancellations involved.}.
We then propose a generalization to 
multiple BFKL exchange in hadron-hadron scattering. 
Starting from these couplings, we then re-derive, for illustration, 
the counting rules presented in the original AGK paper for an arbitrary 
number of exchanged gluons, 
and we apply these rules to single and double inclusive 
jet production cross sections. The comparison with the original AGK paper
allows us to include into our analysis also soft rescattering corrections. 
As a specific example, we consider the double 
BFKL correction to the Mueller-Navelet jet cross section formula \cite{MN}. 

The paper is organized as follows. In section 2 we briefly review 
the logic behind the AGK analysis, and we summarize the results
of ~\cite{BR}. This leads us to define a framework for
studying multiple BFKL 
exchanges in hadron-hadron scattering. In section 3 we discuss the 
counting rules for inclusive cross sections, and in sections 4 and 5 
we turn to single jet and double jet inclusive cross sections. A few details 
are put into an appendix.     

\section{Reggeon unitarity, energy cuts \`a la AGK,\\
and particle-Pomeron couplings in $\gamma^*\gamma^*$ scattering}
\subsection{The nonperturbative AGK rules}

In this section we briefly review the AGK strategy and its application to 
pQCD. We will conclude that the central task is the derivation and the 
study of the coupling of four (or more) reggeized gluons to virtual photons.  

%figure--------- n-poles amplitude ----------------------
\begin{figure}[ht]
\begin{center}
\includegraphics[width=5cm]{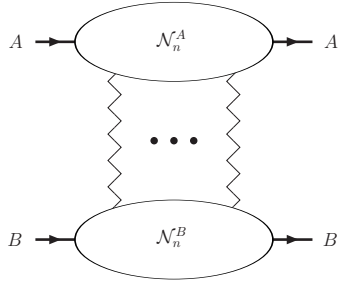}
\caption{Graphical representation of a multi-Regge poles contribution to
the elastic scattering amplitude. The zigzag lines represent pomerons.}
\label{n-poles}
\end{center}
\end{figure}
\begin{figure}[ht]
\begin{center}
\includegraphics[width=15cm]{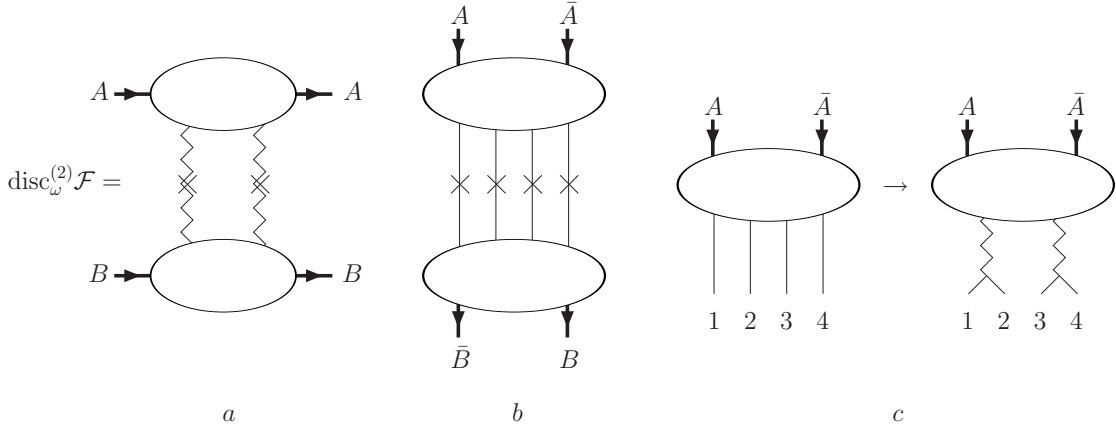}
\caption{Representation of the $t$-channel content of the elastic amplitude:
a) discontinuity across the $2$-poles cut in the $\o$-plane,
b) computation of the coupling $2\rightarrow$4 in the $t$-channel physical
region and
c) isolation of the Regge poles in the (12) and (34) channels.
The crosses in a), b) indicate that the lines are on shell in angular momentum 
and $4$-momentum, resp.}
\label{t-disc}
\end{center}
\end{figure}

%figure--------------------------------------------------
The original AGK paper starts from a multi-Regge pole contribution 
to the elastic scattering amplitude (Fig.\ref{n-poles}), written as a 
Sommerfeld-Watson representation: 
\beq
\cT_{AB}(s,t) =
\int \frac{d \o}{2 i} \xi(\o) s^{1+\o} \cF(\o,t). 
\label{scat}
\eeq
with $\o=J-1$, 
\beq
\label{signature}
\xi(\o)=\frac{\tau-e^{-i \pi \o}}{\sin \pi \o} =
i\frac{e^{-i \frac{\pi}{2} (\omega + \frac{1-\tau}{2})}}
{\cos [\frac{\pi}{2}(\omega + \frac{1-\tau}{2})]}
=i -[\tan{\frac{\pi}{2}\left( \omega+\frac{1-\tau}{2}\right)]},
\eeq
and $\tau = \pm 1$ being the signature. 
The (real-valued) partial wave $\cF(\o,t)$ has singularities in the 
complex $\o$-plane, and the multi-Regge exchange corresponds to a 
particular branch cut. There is a general formula for the discontinuity 
across this cut~\cite{ABSW} (Fig.\ref{t-disc}a):
\beq
\disc_{\o}^{(n)}[\cF(\o,t)] = 2 \pi i 
\int \frac{d\Omega_n}{n!}~\gamma_{\{\beta_j\}}~
\cN_n^{A} (\{\bk_j\}; \o)~\cN_n^{B} (\{\bk_j\}; \o)
~\delta(\o - {\Sigma}_j \beta_j)
\label{regunitarity}
\eeq
\beq
d\Omega_n = (2 \pi)^2\delta^2(\bq - {\Sigma}_j \bk_j)
\prod_{j=1}^n\frac{d^2\bk_j}{(2 \pi)^2}\nonumber
\eeq
where $\bk_j$ ($j=1,...,n$) denotes the transverse momentum of the $j$-th 
Regge pole, $\bq$ is the sum over all transverse momenta 
with $\bq^2 = -t$, and $\alpha(-\bk_j^2)= \alpha_j = 1 + \beta_j$
is the Regge pole trajectory function.
The factor which determines the overall sign has the form
\beq
\gamma_{\{\beta_j\}} = \Im \big[ -i \Pi_j (i\xi_j) \big] =
(-1)^{n-1}\frac{\cos \big[ \frac{\pi}{2} \textrm{$\sum_j$} \Big(\beta_j +
 \frac{1-\tau_j}{2} \Big) \big] }{\textrm{$\prod_j$} \cos \big[\frac{\pi}{2}
\Big( \beta_j + \frac{1-\tau_j}{2} \Big)\big] }
\label{gammafactor}
\eeq
As an example, the contribution of two even-signature Regge poles (Pomerons) 
with intercept close to unity is negative compared to the single pole 
contribution. 
Equation (\ref{regunitarity}) is a `reggeon unitarity equation':
it describes the contribution of the $n$-reggeon $t$-channel state to 
the discontinuity in angular momentum of the partial wave $\cF$
(Fig.\ref{t-disc}).     
In the same way as in a usual unitarity integral particles in the 
intermediate state are to be taken on mass shell, in the reggeon unitarity
integral the reggeons of the intermediate state are on shell in reggeon
energy: as indicated in Fig.\ref{t-disc}c, the Regge pole is a bound state 
of (at least) two particles, and the complex angular momentum of the two 
particles is put equal to the trajectory function of the Regge pole.  
%figure------ t-channel discontinuity -------------------
%figure--------------------------------------------------
The formula \eqref{regunitarity} contains the coupling of $n$ Regge poles to 
the external particles, denoted by $\cN_n({\bk}_j,\omega)$. 
In general, they are functions of $\omega$ and contain, for example poles 
and cuts due to the exchange of Regge poles. This includes, in particular,
the possibility that the reggeons $i$ and $j$ form a composite state.
Depending on the structure of the $\cN_n$, it may be necessary to 
replace, in (\ref{regunitarity}), the $\omega$ -dependent 
$\delta$-function by $\prod_1^n \left(\int d \omega_j \delta(\omega_j - 
\beta_j) \right) \delta (\omega - \sum_j \omega_j)$: in this case, 
the vertex function $\cN_n$ may depend not only upon the total 
angular momentum $\omega$, but also upon the $\omega_j$.
An example will be given futher below.
For these reasons, the vertex functions $\cN_n$ are more general than the 
impact factors.

For later considerations it will be useful to say a few more
words about the origin of this formula.  
Following the idea of Gribov, Pomeranchuk and Ter-Martirosian ~\cite{GPT} 
one starts, in the simplest case, from the four particle intermediate state in the $t$-channel 
unitarity equation in the physical region of the process $A+\bar{A} 
\to B+\bar{B}$ (Fig.\ref{t-disc}b).
Above the 4-particle state, the $2 \to 4$ amplitude 
for the process $A+\bar{A} \to 1+2+3+4$ appears. 
This scattering amplitude (and the corresponding one below the 
4-particle state) is expanded in terms of  
partial waves, inserted into the $t$-channel unitarity 
equation. Defining analytic continuations in angular momentum and helicity 
variables, using Sommerfeld-Watson transformations, and performing 
parts of the $t$-channel unitarity phase space integrals, one moves from the 
physical $t$-channel region to negative $t$-values. Retaining Regge poles 
in the $(12)$ and $(34)$ channels (illustrated in Fig.\ref{t-disc}c), 
one finally arrives at the discontinuity 
of the $t$-channel partial wave across the two-reggeon cut, given by 
\eqref{regunitarity}. This form of 
$t$-channel unitarity is called 'reggeon unitarity':
in the Regge description of high energy scattering processes
the partial waves satisfy reggeon unitarity relations, and the single 
pole, double, triple... exchanges can be isolated by computing 
discontinuities across the corresponding cuts in the 
angular momentum plane. The analytic form of the discontinuity equation  
given in \eqref{regunitarity} is universal,
whereas the reggeon-particle couplings entering the equation, $\cN_n$, have 
to be computed from the underlying theory. 

\begin{figure}[ht]
\begin{center}
\includegraphics[width=10cm]{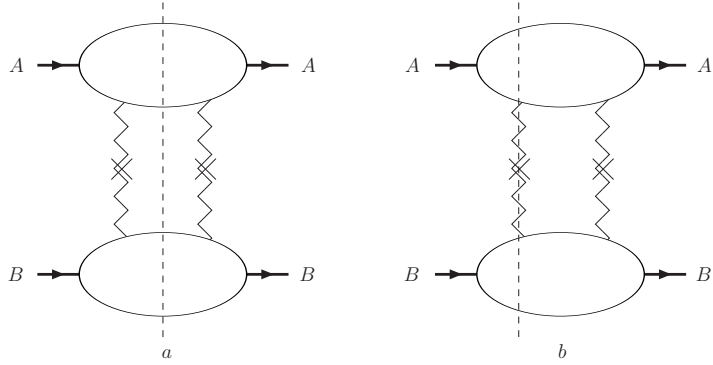}
\caption{Some examples of the different $s$-channel cuts of the $2$-pomeron
contribution: a) diffractive and b) multiperipheral intermidiate states.}
\label{2pcuts}
\end{center}
\end{figure}

The central goal of the AGK analysis is the decomposition of the n-reggeon 
exchange contribution in terms of $s$-channel intermediate states. To be 
precise, one is interested in the total cross section, i.e. in 
the absorptive part or, equivalently, in the discontinuity w.r.t. energy 
of the scattering amplitude \eqref{scat}. 
It is quite obvious that the  
absorptive part of the amplitude (1) will consist of several different 
contributions: each piece belongs to an energy cut line across 
Fig.\ref{n-poles}, and there are several different ways of drawing 
such energy-cutting lines. Each of them
belongs to a particular set of $s$-channel intermediate states. 
For example, a cutting line between reggeons (Fig.\ref{2pcuts}a)
belongs to double diffractive production on both sides of the cut:
there is a rapidity gap between what is inside the upper blob and the 
lower blob. When relating 
this contribution with the full diagram in
Fig.\ref{n-poles} one requires a `cut version' of the reggeon
particle couplings
$\cN_n$. Similarly, the cut through a reggeon (Fig.\ref{2pcuts}b)
corresponds to a so-called 
multiperipheral intermediate state, and another cut version of $\cN_n$ 
appears. The basis of the AGK analysis is the observation that, under very 
general assumptions for the underlying dynamical theory, 
the couplings $\cN_n$ are fully symmetric under the exchange 
of reggeons, and all their cut versions are identical. 
This property then allows to find simple relations between 
the different cut contributions, and to derive a set of counting rules. 
%figure---------- 2-pomeron cuts -----------------------
%figure--------------------------------------------------
 
\subsection{The pQCD case}
\begin{figure}[ht]
\begin{center}
\includegraphics[width=5cm]{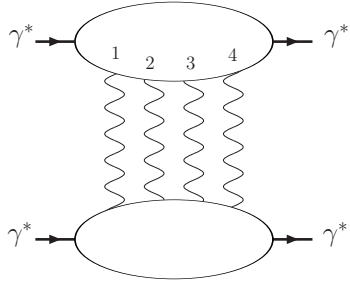}
\caption{Exchange of four reggeized gluons in pQCD. The wavy lines represent
reggeized gluons. The blobs above and below denote the functions 
$\cN_4(\{\bk_j\},\omega)$, computed in pQCD.}
\label{gamma-gamma}
\end{center}
\end{figure}

\begin{figure}[ht]
\begin{center}
\includegraphics[width=15cm]{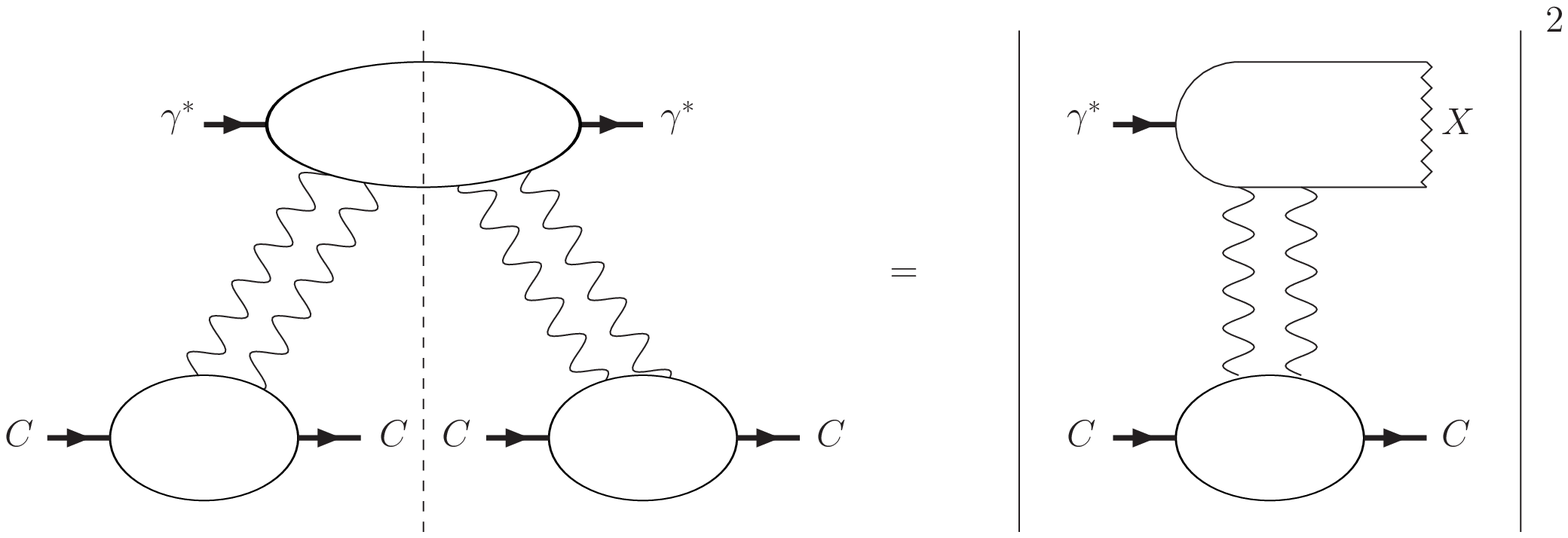}
\caption{Six point amplitude for the computation of the high-mass diffraction
cross section in the triple Regge limit.}
\label{triplereg}
\end{center}
\end{figure}

From this brief review it follows that the central task of 
performing the AGK analysis in pQCD requires the computation and study of 
the coupling functions $\cN_n$. The simplest task is the study of the 
two-Pomeron exchange. Since the BFKL Pomeron 
is a composite state of two reggeized gluons, we have to start from the 
exchange of four reggeized gluons (Fig.\ref{gamma-gamma}).
The blobs above and below denote the couplings $\cN$, and they 
will be discussed below. They have a nontrivial content of reggeized gluons, 
and, in particular, any pair of gluons $1,2,...,4$ can come from a composite 
BFKL state contained in $\cN$.   This diagram has to be compared 
with Fig.\ref{n-poles}: the elementary Regge poles in pQCD are 
the reggeized gluons, and the Pomeron appears as a composite state.    
%figure----- gamma-gamma to 4 reggeized gluons ---------
%figure--------------------------------------------------
Rather than going in all detail through    
the chain of arguments described before (which would lead us to the 
study of a $t$-channel $2 \to 8$ process!), we mention a 'shortcut' path 
which takes us, in an easier way, to the desired coupling of four reggeized 
gluons to external particles. It is based upon the observation that the 
same coupling which 
appears in the reggeon unitarity equation (\ref{regunitarity}) for the 
discontinuity across the four reggeon cut is also contained in the diffractive cross section formula 
for low and high-mass diffraction (triple Regge limit): the process 
(Fig.\ref{triplereg}) $\gamma^*+C \to X +C$ where $X$ 
sums over low and high mass diffractive states (with squared mass  
$M_X^2$) of the incoming projectile $\gamma^*$. 
%figure------------ triple Regge limit ------------------
%figure--------------------------------------------------
In the limit of large $M_X$ and large energy $s$ with the restriction 
$M_X^2 \ll s$, this process is dominated by Pomeron exchanges in the 
lower $t$ channels which in pQCD are BFKL ladders. By cutting 
these $t$-channels (more precisely: by taking cuts across the angular 
momentum variables in the lower $t$-channels), we arrive at the coupling 
of four reggeized gluons to the external particle $\gamma^*$.     
This is the method which was used in ~\cite{BW} for deriving the 
coupling of four reggeized gluons to the virtual photon, named $\cD_4$. 
It is defined to contain the energy denominator $1/(\omega - \sum_{j=1}^{4})$ 
for the four reggeized gluons; its amputated counterpart, 
$C_4 = D_4 (\omega - \sum_{j=1}^{4})$ will lead to a pQCD model for the 
coupling function $\cN_4$. The study of $D_4$, therefore, 
provides the starting point for the AGK analysis in pQCD. 
In the following we will discuss the function $D_4$, keeping in mind 
that at the end we have to multiply by $(\omega - \sum_{j=1}^{4})$.
Before inserting this coupling into the discontinuity formula, 
we have to address the complications arising from the color degree of freedom 
and from the reggeization of the gluon. 
      
\begin{figure}[ht]
\begin{center}
\includegraphics[width=15cm]{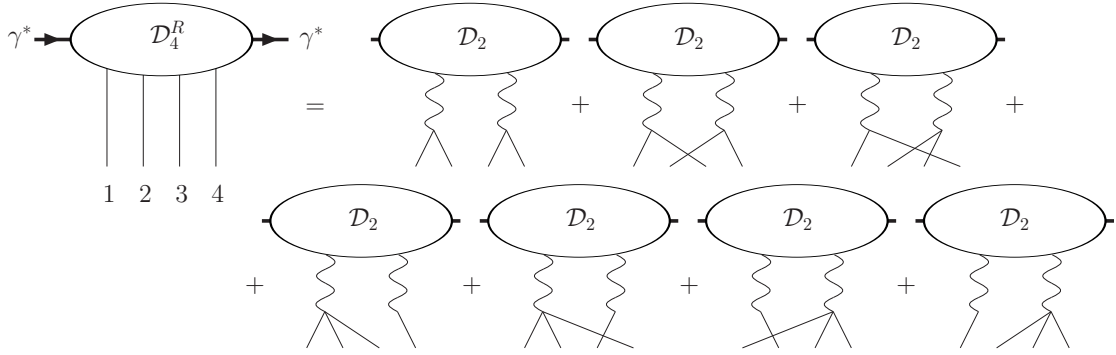}
\caption{Decomposition of the reggeizing part of the four reggeon amplitude
$\cD_4^R$ in term of reggeized gluons.}
\label{D4R}
\end{center}
\end{figure}

Let us recapitulate the main properties of $\cD_4$. 
From the $t$-channel point of view it is natural to demand that the 
reggeon-particle coupling 
that enters the reggeon unitarity equation satisfies certain symmetry 
requirements. In the simplest case, it should be symmetric under 
the exchange of the reggeons, i.e. their momenta and color labels.
On the other hand, we know from the BFKL equation that there are two 
different $t$-channel states. The Pomeron state which belongs to the color 
singlet is completely symmetric; on the other hand, when the BFKL 
amplitude is projected onto the antisymmetric color octet, it satifies the 
bootstrap equation, and the two gluons 
'collapse' into one single gluon.  It is therefore not unexpected that also 
the four reggeon amplitude $\cD_4$ contains antisymmetric configurations which
satisfy bootstrap equations. One of the main results of the analysis 
of ~\cite{BW} is the complete decomposition of $\cD_4$ into two pieces:
\beq
\cD_4=\cD_4^I + \cD_4^R
\label{D4decomp}
\eeq
Here the first term, $\cD_4^I$, is completely symmetric under the exchange of 
any two gluons, 
whereas the second one, $\cD_4^R$, is a sum of antisymmetric terms which,
as a result of bootstrap properties, can be expressed in terms of 
two-gluon amplitudes, $\cD_2$. In a graphical way, $\cD_4^R$ is illustrated in 
Fig.\ref{D4R}:
after making use of the bootstrap properties, $\cD_4^R$ is reduced to 
a sum of $\cD_2$ functions. Under the exchange of the two reggeized gluons,
$\cD_2$ is symmetric. It is only after this decomposition has been 
performed, and we have arrived at reggeon particle couplings with `good' 
properties, that we can start with the AGK analysis. 
%figure------------------- D4R --------------------------
%figure--------------------------------------------------

The construction of $\cD_4^I$ implies that the same function appears when,
in the AGK analysis, one computes different energy discontinuities of 
$\gamma^*$-$\gamma^*$ scattering, 
no matter where the cutting line runs, between gluon `1' and `2', 
or between `2' and `3', or between `3' and `4'. 
Since in pQCD the Pomeron is a bound state       
of two gluons, the two-Pomeron state is formed, e.g., by the pairs $(12)$ 
and $(34)$. Thus cuts through the Pomeron or between 
two Pomerons all lead to the same two-Pomeron particle vertex. 
This means that $\cD_4^I$ satifies all the requirements listed in the 
AGK paper. 

Starting from these functions $\cD_4^I$ and $\cD_4^R$, the investigation in 
~\cite{BR} has shown in some detail how the AGK counting rules work in 
pQCD: the analysis has to be done 
seperately for $\cD_4^I$ and $\cD_4^R$. For the first piece, we obtain  
the counting arguments for the Pomerons (which is even-signatured) 
given by AGK; 
here the essential ingredient is the complete symmetry of $\cD_4^I$ under the 
permutation of reggeized 
gluons. In the latter piece, the odd-signature reggeizing gluons lead to 
counting rules which are slightly different from those of the even signature 
Pomeron: once the bootstrap properties have been invoked and $\cD_4^R$ is 
expressed in terms of $\cD_2$ functions, cutting lines through the reggeized 
gluon appear. Since it carries negative signature, the relative weight 
between cut and uncut reggeon is different from the Pomeron. It should be    
stressed, however, that the generalization of the AGK analysis 
to odd signature reggeons is contained in the 
AGK paper; in pre-QCD times, however, there was no 
obvious reason for considering Regge poles other than the Pomeron.
        
%figure--------------------------------------------------
\begin{figure}[ht]
\begin{center}
\includegraphics[height=3cm]{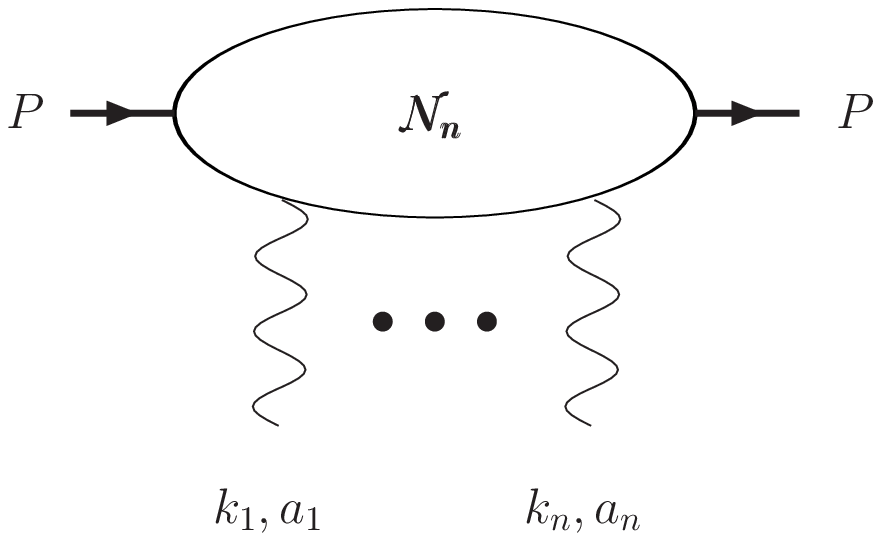}
\caption{Graphical illustration of the coupling
$\cN_n(\bk_1,a_1;\ldots;\bk_n,a_n;\o)$ of a proton to $n$ reggeized gluons.}
\label{Nn}
\end{center}
\end{figure}
%figure--------------------------------------------------

Let us now turn to the main goal of this paper, to the use of the pQCD 
cutting rules in a nonperturbative environment (e.g.multi-ladder exchanges 
in $pp$ scattering). Basic ingredients are the 
nonperturbative couplings of $n$ reggeized gluons to the proton.
In order to justify the use of pQCD we need a hard scattering subprocess:  
we will assume that all reggeized gluons are connected to some hard 
scattering subprocess; consequently, each gluon line will have its 
transverse momentum in the kinematic region where the use of pQCD can be 
justified. Since AGK applies to the high energy limit 
(i.e. the small-$x$ region), all $t$-channel gluons are reggeized.    
Based upon the analysis in pQCD,
we formulate a few general conditions which the nonperturbative        
couplings have to satisfy. 
Following the original AGK paper, we will denote these couplings      
by $\cN_n(\bk_1,a_1;\ldots;\bk_n,a_n;\o)$,
where $\bk_i$ and $a_i$ refer to transverse momenta 
and color of the $i$th gluon, and require that
\begin{quote} \em
(i) they are symmetric under the simultaneous exchange of momenta 
and color, e.g. $(k_i,a_i)\leftrightarrow (k_j,a_j)$;\\
(ii) cut and uncut vertices are identical, independently where the 
cut line enters.
\end{quote}
In the following sections we will work out a few results which follow from 
these conditions.  

We conclude this section with a remark on the reggeizing pieces, $D_4^R$. 
The analysis of 
$\cD_2$, $\cD_3$, and $\cD_4$ in ~\cite{BW},
and of $\cD_5$ and $\cD_6$ in ~\cite{BE}
lead to the following conjecture about the general coupling of $n$ 
reggeized gluons to virtual photons. Functions with an even number of lines, 
$\cD_{2n}$, contain
\begin{itemize}
\item a piece which is completely symmetric, $\cD_{2n}^I$;
\item a second piece which can be expressed in terms of $\cD_{2n-2}$, 
by either having two of the outgoing reggeized gluons each split into two gluons 
or one of the outgoing gluons split into three gluons; 
\item a third piece which is reduced to $\cD_{2n-4}$ with further splittings 
at the lower end etc. 
\end{itemize}
Finally, functions with an odd number, $\cD_{2n+1}$,
\begin{itemize}
\item  can always be reduced to $\cD_{2n}$, with one splitting at the lower 
end.
\end{itemize}
As a consequence, also these reggeizing pieces can be reduced to the 
couplings $\cN_{2n}$.
Since the reggeization of the gluon is a general 
feature of QCD, valid to all orders in perturbation theory, the appearance 
of these reggeizing pieces can be expected to happen also in a 
nonperturbative coupling; in other words, in addition to the functions 
$\cN_{2n}$ mentioned before, we expect also contributions with the same 
functions $\cN_{2n}$, but one (or more) of the reggeized gluons split into 
two gluons etc.  

In the following we will apply the AGK analysis to $pp$ scattering,
assuming that the couplings of $n$ gluons to the proton satisfy the 
requirements listed above. In this paper we will not analyse yet
the reggeizing pieces; their analysis will be taken up in a future 
paper. Another application of the AGK analysis is DIS, in particular 
the production of diffractive final states. Such an analysis cannot avoid to 
include the reggeizing pieces: to lowest order in $\alpha_s$ 
the coupling of $n$ gluons to the virtual photon consists of reggeizing 
pieces only and hence cannot be 
ignored.  

%Recently the case of one and two gluon inclusive production has started to be
%analysed~\cite{Kovchegov1_2, Braun2005} in relation to the dipole picture.
%There is still a debate and some contraddictory results but the
%emerging picture is that there may be a contribution from the transition
%vertices (cutted) which originate a jet to the counting related to AGK rules
%(a generaled version of).
%Nevertheless any kind of extra contributions seems subleading in the high
%energy limit. We shall therefore ignore them in our analysis.
   
%%%%%%%%%%%%%%%%%%%%%%%%% Inclusive C.S. %%%%%%%%%%%%%%%%%%%%%%%%%
\section{Inclusive cross section}
In this section we discuss the AGK rules for inclusive cross sections.
Since we are not asking for a final state which provides a hard scale,
such a discussion might seem somewhat academic: rescattering in $pp$ 
collisions is described by multiple exchanges of nonperturbative Pomerons, not 
simply of bound states of reggeized gluons. For consistency, however, it is 
important to show that BFKL exchanges lead to the same counting rules as the 
nonpertubative Pomerons. Also, for the modelling of events with multiple 
scattering events, it will be useful to illustrate the pattern of 
AGK cancellations.       

\subsection{The nonperturbative case}

Let us begin by recapitulating the counting rules of the original AGK paper. 
The contribution of the $n$ Pomeron cut to the scattering amplitude
is given by
\beq
\cT_{AB}^{\textrm{n-cut}}(s,t) = 
\int \frac{d \o}{2 i} \xi(\o) s^{1+\o}~
\disc_{\o}^{(n)}[\cF(\o,t)], 
\label{ampdecomposition}
\eeq
and the discontinuity of the partial wave has the form given in 
(\ref{regunitarity}). Doing the $\omega$ integral we arrive at:   
\bea
\cT_{AB}^{\textrm{n-cut}}(s,t) &=&
\pi \int \frac{d\Omega_n}{n!} \gamma_{\{\beta_j\}}~
 \xi(\beta)s^{1+\beta}~
\cN_n^{A}~
\cN_n^{B}
\label{ampexp}
\eea
where we have defined $\beta \equiv \beta(\{\bk_j\}) = \sum_{j=1}^n \beta_j$
and $\cN_n^{A,B} \equiv \cN_n^{A,B}(\{\bk_j\}; \beta )$.
The energy discontinuity (imaginary part) is obtained by simply 
replacing the signature factor (\ref{signature}) by $1$. 
As we have explained before, the aim of AGK is to obtain the same result for 
the discontinuity from the unitarity equation, i.e. from products of 
production amplitudes. Each such contribution can graphically be illustrated 
by a cutting line (Fig.3). One of the main results of AGK states that
the $s$-discontinuity of the $n$-cut contribution to the amplitude,
$\disc_s[\cT_{AB}^{\textrm{n-cut}}(s,t)]$,
can be written as a sum over the number
$k$ of cut pomerons ($k=1,...,n$),
\beq
\cA^n(s,t) \stackrel{\textrm{def}}{=}
\disc_s[\cT_{AB}^{\textrm{n-cut}}(s,t)] = \sum_{k=0}^n \cA_k^n(s,t),
\label{muldecomp}
\eeq
and the terms in the sum are
\beq
\cA_k^n(s,t) = 2 \pi i \int \frac{d\Omega_n}{n!}~ \cF_k^n~
s^{1+\beta}~ \cN_n^{A}~ \cN_n^{B},
\eeq
where we have introduced the AGK factors
\beq
\cF_k^n =
\Bigg\{ \begin{array}{ll}
(-1)^n~2^{n-1}+\gamma_{\{\beta_j\}}       & \textrm{if $k=0$} \\
(-1)^{n-k}~2^{n-1}\binom{n}{k}            & \textrm{if $k>0$}
\end{array}.
\label{agk}
\eeq
If a (nonperturbative) pomeron is viewed as a multiperipheral 
chain of secondary particles, the cut of each pomeron
gives a uniform distribution in rapidity, and the sum in (\ref{muldecomp})
leads to density fluctuations. 

The simplest case, the two-Pomeron 
exchange, has the three contributions illustrated in Fig.8: 
%figure-------irreducible diagrams------------------------
\begin{figure}[ht]
\begin{center}
\includegraphics[width=15cm]{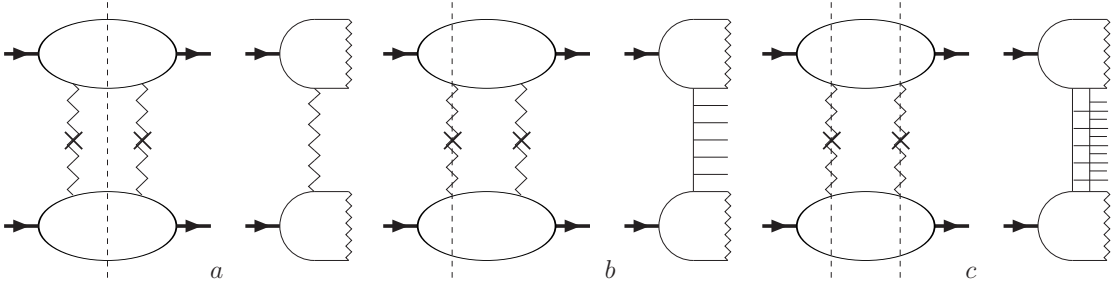}
\caption{There are three different ways to cut the two-pomerons diagram:
a) diffractive cut, b) single multiplicity cut and c) double multiplicity cut.}
\label{fig:agk2p}
\end{center}
\end{figure}
%figure--------------------------------------------------
Figure \ref{fig:agk2p}.a shows the diffractive cut: all the pomerons are left
uncut, and there is a rapidity gap between the fragmentation regions of the
two particles. Figure \ref{fig:agk2p}.b represents a single multiplicity
cut in which only one pomeron has been cut. Eventually, the situation
shown in figure \ref{fig:agk2p}.c corresponds to two cut pomerons, and the
multiplicity of particles is doubled with respect to the previous case.
Neglecting the real part of the pomeron signature factor \eqref{signature}
it reduces to the imaginary unit $i$, and the $\gamma$ factor
\eqref{gammafactor} is just $(-1)^{n-1}$.
From \eqref{agk} we obtain for the weight factors the well-known 
results:
\bea
1:            & \textrm{diffractive}         & (k=0) \nonumber \\
-4:           & \textrm{single multiplicity} & (k=1) \\
2:            & \textrm{double multiplicity} & (k=2).\nonumber
\label{eq:agk4}
\eea
In other words, the different contributions are in the proportion
\beq
\cA^2_0:\cA^2_1:\cA^2_2 = 1:-4:2.
\label{prop}
\eeq
This result can be summarized as follows: final states with a rapidity 
gap ($k=0$) are accompanied by other final states with double density ($k=2$),
and their respective cross sections come with the 
relative weight given by \eqref{prop}. 
At the same time, these final states are connected with corrections 
to the cross section of final states with normal density ($k=1$), 
as contained in \eqref{prop}. 

For later use we note an important generalization: suppose we
substitute one of the soft pomerons by a different Regge pole, and we
concentrate on those contributions where this Regge pole is cut 
(later on, in the context of inclusive jet production, we shall apply this 
argument to a hard cut gluon ladder). Following the previous argument
used to derive the usual AGK rules, we simply sum over 
cut and uncut soft pomerons, where the uncut soft Pomerons appear on both the
left and the right side of the cut reggeon. As a result, 
the $\gamma$ factor in the first line of \eqref{agk} does not appear, and
for the contribution of $k$ cut Pomerons we simply obtain:
\beq
\cF^n_k \propto (-1)^{n-1-k} \binom{n-1}{k},~~~~~~~k=0,...,n-1
\eeq
Clearly the sum over $k$ vanishes identically: soft multi-Pomeron 
corrections to a single cut Regge pole cancel. 
This argument is easily generalized to two or more singled out cut 
Regge poles.  

So far we have discussed the AGK counting only at one rapidity value.
The AGK paper also addresses the question of how to continue 
the $s$-discontinuity cutting lines inside the upper or lower vertex 
function $\cN_n$. When following, for example, the cut Pomerons 
inside $\cN_n$ (or cutting lines running between Pomerons) 
one faces the question of how these cutting lines pass through 
Pomeron interaction vertices ('cut vertices').
AGK constraints can be formulated only for a very restricted class 
of interaction vertices (in particular, for the $1 \to n$ Pomeron vertex). For 
the general case (for example, for the $2 \to 2$ vertex) this is not the case;
only explicit models, e.g. calculations in pQCD, can provide further 
information.

\subsection{The pQCD case}
Let us now derive the counting rule \eqref{agk} in pQCD.
Since a Pomeron appears as a bound state of two reggeized gluons,
the discussion of $n$ Pomerons has to start from $2n$ reggeized gluons.
Moreover, in LLA the signature factor of the reggeized gluon is real,
$\xi_\mathds{G}(\bk) = -2/\pi \bk^2$; therefore, working in LLA, the
diagrams with any cut reggeized gluon are suppressed and will not be 
considered.
The derivation of \eqref{agk} follows from straightforward combinatorics.  
The simplest case, $n=2$, has been discussed in ~\cite{BR}, and we 
can use these results (Fig.9) for illustrating the general proof.   

%figure-------irreducible diagrams------------------------
\begin{figure}[ht]
\begin{center}
\includegraphics[width=8cm]{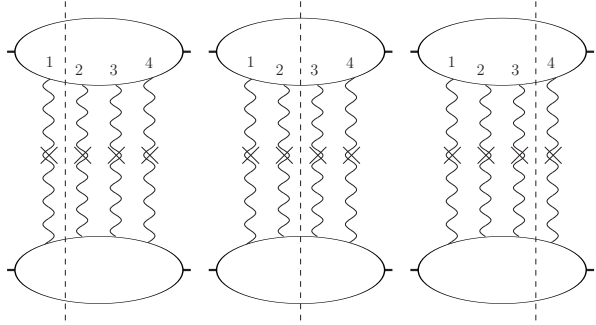}
\caption{The three $s$-channel cuts contributing to the four
reggeized gluon exchange.}
\label{fig:agk4}
\end{center}
\end{figure}
%figure--------------------------------------------------

The situation is depicted in figure \ref{fig:agk4}: each term denotes 
the product of production amplitudes, and it is understood that 
we sum and integrate over the produced gluons. We again take the 
discontinuity in $\omega$ across the four $t$-channel gluons 
(denoted by crosses),
and all the produced gluons are absorbed inside the blobs above and below.
In the first diagram, the production amplitude on the left of the cutting line 
(vertical dashed line) contains the exchange of one gluon, 
the amplitude on the rhs three gluons.
Three gluon exchange comes with a negative sign (from the $\gamma$-factor 
in (\ref{regunitarity}) and from the signature factor 
in (\ref{scat})); moreover, there is a symmetry factor $1/3!$.
Similarly, the second term in Fig.\ref{fig:agk4} denotes the square of two 
production amplitudes: two-gluon exchange is purely imaginary and has the 
symmetry factor $1/2!$. Note that, in contrast to the 2 pomeron exchange 
discussed above, in the case of reggeized gluons we do not need to consider
cutting lines inside the reggeized gluons: compared to an uncut gluon, 
a cut gluon line is suppressed in order $\alpha_s$. 
               
Since we are looking for the contributions of BFKL Pomerons which are 
bound states of two
gluons, we must consider all possible pairings among the reggeized gluons;
using the notation $(i_1j_1)(i_2j_2)$ to indicate that gluon $i_1$
forms a bound state with the gluon $j_1$, and gluon $i_2$ with $j_2$, 
the three possibilities are: $(12)(34)$, $(13)(24)$ and $(14)(23)$.\\
Let us first consider the `diagonal' configuration, in which the 
pairings in the upper and in the lower blob are identical.   
One easily sees that these different configurations
contribute with different multiplicities.
Starting with the first graph in Fig.\ref{fig:agk4}, in all three 
possible pairings one of the gluon pairs is cut, and the other one is not. 
This means that all the three configurations contribute
with single multiplicity ($k=1$); the weight factor is $- 3 \times 1/3!$.
The same argument holds for the third graph in Fig.\ref{fig:agk4}. 
In the second graph, the configuration $(12)(34)$ does not have any cut pair;
this contributes to the diffractive term ($k=0$), and the weight is 
$(1/2!) \times (1/2!) = 1/4$. The other two pairings have
both gluon pairs cut, and therefore they contribute to the
double multiplicity term ($k=2$); the weight factor is 
$2 \times (1/2!) \times (1/2!) = 1/2$. This agrees with the AGK result
(\ref{prop}).

It should be clear how to generalize this counting to the exchange of $2n$ 
reggeized gluons. The general result for the 
contribution of $k$ cut Pomerons assumes the form 
(for the explicit computation see Appendix A):
\beq
\tA_k^n(s,t) = 2 \pi i~ \tilde{\cF}_k^n
\int \frac{d\Omega_{2n}}{n!}~ s^{1+\tilde{\beta_n}}~
\cN_{2n}~ \cN_{2n}~
\tilde{\gamma}_{\{\bk_j\}},
\label{agkpQCD}
\eeq
where
\beq
\tilde{\cF}_k^n =
\Bigg\{ \begin{array}{ll}
(-1)^n~2^{n-1}+(-1)^{n-1}             & \textrm{if $k=0$} \\
(-1)^{n-k}~2^{n-1}\binom{n}{k}        & \textrm{if $k>0$}
\end{array}.
\label{tildeF}
\eeq
and
\beq
\tilde{\gamma}_{\{\bk_j\}} \stackrel{\textrm{def}}{=}
\frac{2^{1-n}}{\pi} \prod_{j=1}^{2n} \xi_\mathds{G}(\bk_j) =
\frac{2^{1-n}}{\pi} \prod_{j=1}^{2n} \frac{2}{\pi \bk_j^2}.
\label{pQCDgamma}
\eeq
Here we use the tilde symbol to indicate that the result is obtained in pQCD.
Note that the weight factors $\tilde{\cF}^n_k$ coincide with those defined
in \eqref{agk} for $k\ne0$, while for $k=0$ they coincide when the
real part of the soft pomeron signature factor is neglected
($\gamma_{\{\beta_j\}}=(-1)^{n-1}$).

It should be stressed that this discussion has made use only of the
very general symmetry properties of the $4$-gluon vertex functions above 
and below the four gluon state. Each vertex function can very well have a 
complicated internal structure of reggeized gluons; for example, it  
may consist of a single BFKL ladder which splits into four gluons.
In this case, our analysis has demonstrated the AGK rules inside a closed 
Pomeron loop. The crucial ingredient is the symmetry structure, and in this 
example it is garanteed by the form of the $2\to4$ gluon vertex. 
In an anologous way one can generalize the 
validity of the AGK analysis to general reggeon diagrams in pQCD. 

What remains is the discussion \cite{BR} of the case where, 
in Fig.\ref{fig:agk4}, the pairings above and below the four gluon state 
do not match (non-diagonal configurations): an example is given in
Fig.\ref{fig:4reg-nondiag}a. The corresponding final states are depictured in
Fig.\ref{fig:4reg-nondiag}b. Here the situation is the following:
in the upper rapidity interval we have multiplicity $k$ which can take 
one of the values $0$, $1$, or $2$ ($k=0$ in our 
example). Each multiplicity $k$ comes with a relative weight, described by 
the three-component vector $(1,-4,2)$. In the lower rapidity interval 
we have multiplicity $k'$ (in our example $k'=2$). 
Contributions to this $k'$ can come from different 
$k$ in the upper interval, so the transition from the upper to the lower 
interval - named a 'switch' - can be described by a $3 \times 3$ 
matrix, whose elements are defined to the fraction of configurations 
that lead from $k$ to $k'$:   
%figure-------irreducible diagrams-------- ----------------
\begin{figure}[ht]
\begin{center}
\includegraphics[width=6cm]{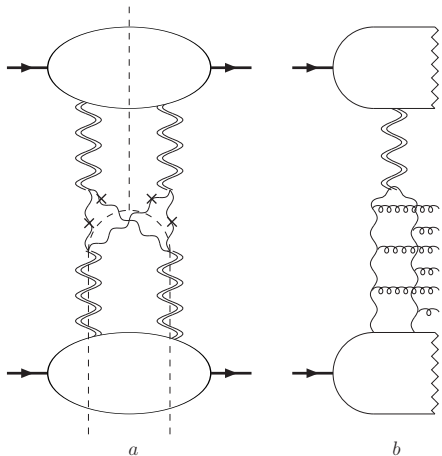}
\caption{An example of non-diagonal configuration: a double wavy line
represent a two reggeized gluons bound state (they can be freely moved
due to the symmetry of the impact factors). The bound state configuration
is not the same above and below the cut. In a) the cut forward amplitude
is shown, while in b) we depict an example of a final state coming from the cut.}
\label{fig:4reg-nondiag}
\end{center}
\end{figure}
%figure--------------------------------------------------
\bea
\cM = \left( \begin{array}{ccc}
             0 & 0 & \frac{1}{2}\\
             0 & 1 & 0 \\
             1 & 0 & \frac{1}{2}
             \end{array}
\right)
\eea
(here $k$ labels the columns, $k'$ the rows). Now one easily verifies that the
vector $\mbf{\cF}^t = (1, -4, 2)$ is eigenvector of this matrix $\cM$: 
this implies that the ratios: diffraction : single density : double density 
holds for both 
the upper and the lower rapidity interval, i.e. it is invariant.     

This pattern holds for an arbitrary number, $2n$, of exchanged gluons. 
The number $k$ of cut BFKL Pomerons 
denotes the density of gluons in the final state, and for diagonal 
configurations the AGK factors given in \eqref{tildeF} measure 
the relative weight of states with density $k$. If we move to the neighbouring 
rapidity interval, the density changes to $k'$: the transitions from 
$k$ to $k'$ define a matrix $\cal M$, and the eigenvectors of this 
matrix are formed by the AGK factors in \eqref{tildeF}. 
They define, for each rapidity interval, the relative weight of density $k$. 
Some details are given in the Appendix.
%%%%%%%%%%%%%%%%%%%%%%%%% Eikonal part %%%%%%%%%%%%%%%%%%%%%%%%%%%%%%%
\subsection{An example}
More detailed results can be derived if one assumes a specific model 
for the couplings $\cN_{2n}(k_1,a_1;\ldots;k_{2n},a_{2n};\omega)$ to 
hadrons $A$ and $B$. A popular 
choice is the eikonal model. In order to satisfy our symmetry requirements,
we have to start from the ansatz:
\bea
\cN_{2n}^A(\bk_1,a_1;\ldots;\bk_{2n},a_{2n};\omega)=
~~~~~~~~~~~~~~~~~~~~~~~~~~~~~~~~~~~~~~~~~~~~~~~~~~~~~~~~~~~~~~~~~~~~~~~
\nonumber\\
\frac{1}{\sqrt{(N_c^2-1)^n}}
\Biggl( \phi^A(\bk_1,\bk_2;\omega_{12})
\delta_{a_1a_2}
\cdot
... \cdot\phi^A(\bk_{2n-1},\bk_{2n};\omega_{{2n-1},{2n}})
\delta_{a_{2n-1}a_{2n}}
+\sum_{Permutations}
\Biggr)\;.
\eea
Here $\omega_{12}= \omega_1 + \omega_2$, and each factor 
$\phi^A(k_1,k_2;\omega_{12})\delta_{a_1a_2}$ represents a BFKL amplitude 
in the color singlet state, convoluted with an impact factor 
$\phi_0^A(\bk_1,\bk_2)$. 
The sum has to extend over all possible 
pairings of the gluons $1,...,2n$: $(i_1,j_1),...,(i_n,j_n)$
(alltogether, there are $C^n = (2n-1)!!$ such possibilities). When inserting 
this ansatz for $\cN^A_{2n}$ and $\cN^B_{2n}$ into (\ref{discspQCD}), 
(\ref{discspQCD1}) and performing the counting described in the first 
subsection of the Appendix we proceed in the following way:\\
(a) we rewrite (\ref{discspQCD}) in the same way as (\ref{regunitarity}), 
i.e. we define the Sommerfeld-Watson transformation $\tF^n(\o,\bq)$
of the scattering amplitude $\tT^n(s,\bq)$ 
and take the discontinuity across the $2n$-reggeon cut:
\beq
\disc_{\o} \tF^n(\o,\bq) =
4 i \sum_{j=1}^{2n-1} S^n_j~
\int d\Omega_{2n}~ \cN^A_{2n}~ \delta(\o-\Sigma_j \beta_j)~ \cN_{2n}^B~
\prod_j \xi_\mathds{G}(\bk_j)\;.
\eeq 
(b) when combining the $\cN_{2n}$ from 
above and from below, we retain only the diagonal combinations, 
i.e. those combinations where the pairings above and below match.
In this way we sum over all bound states. 
Formally, this coincides with the large-$N_c$ limit.
Each term obtained in this way is a product of $n$ BFKL exchanges.
\\
(c) in each of these products of $n$ BFKL exchanges we have to apply the 
discussion given after eq.(\ref{gammafactor}): 
for each pair of gluons $(ij)$ we define the variable 
$\omega_{ij}$ and we substitute  
$\delta(\omega - \sum \beta_j) \to [\prod_{(ij)} \int d\omega_{ij}
\delta(\omega_{ij} - \beta_i - \beta_j)] \delta(\omega -
\sum \omega_{ij})$. In order to take care of momentum conservation we 
write the measure as
\beq
d\Omega_{2n} = \int d^2\bb e^{i \bb \cdot \bq}
\prod_{j=1}^{2n}\frac{d^2\bk_j}{(2 \pi)^2}e^{-i \bb \cdot \bk_j }.
\eeq
The discontinuity then assumes the form:
\bea
\disc_\o \tF^n(\o,t) = 
4 i \sum_{j=1}^{2n-1} S^n_j C^n~
\int d^2\bb e^{i \bb \cdot \bq}
\prod_{(ij)} \Bigg[
\int \frac{d^2\bk_{i}}{(2 \pi)^2}
\int \frac{d^2\bk_{j}}{(2 \pi)^2}
e^{-i \bb \cdot (\bk_{i}+\bk_{j})} \nonumber \\
\label{discFna}
\cdot \int d\o_{ij}
\xi_\mathds{G}(\bk_{i}) \xi_\mathds{G}(\bk_{j})
\phi^A(\bk_{i},\bk_{j},\o_{ij})
\delta(\o_{ij} -\beta_{i}-\beta_{j})
\phi^B(\bk_{i},\bk_{j},\o_{ij})
\Bigg]
\delta(\o - \Sigma_{(ij)} \o_{ij} )\;.
\eea
(d)
Defining the variable
\bea
\bq_{ij} = \bk_{i} + \bk_{j},
\eea
using the result (\ref{cutcounting}) of the Appendix
and observing that, since $\xi_\mathds{G}(\bk)$ is real,
$\xi_\mathds{G}(\bk_{i})\xi_\mathds{G}(\bk_{j})=
\Im[-i(i\xi_\mathds{G}(\bk_{i}))(i\xi_\mathds{G}(\bk_{j}))]=
\gamma_2(\bk_{i},\bk_{j})$,
we can write the multiplicity k (for $k > 0$) contribution to
\eqref{discFna} in the following form:
$$
\disc_\o \tF^n_k(\o,t) =
4 i \frac{(-1)^{n-k}}{k!(n-k)!}
\int d^2\bb e^{i \bb \cdot \bq}
\prod_{(ij)} \Bigg[
\int \frac{d^2\bq_{ij}}{(2 \pi)^2} ~ e^{-i \bb \cdot \bq_{ij}}
\int d\o_{ij}\;  \cdot
$$ 
\beq
\label{discFnb}
\int \frac{d^2 \bk_i}{(2 \pi)^2}
\gamma_2(\bk_i,\bk_j)
\phi^A(\bk_{i},\bq_{ij}-\bk_{i},\o_{ij})
\delta(\o_{ij}-\beta_i - \beta_j)
\phi^B(\bk_i,\bq_{ij}-\bk_i,\o_{ij})
\Bigg]
\delta(\o - \Sigma_{(ij)} \o_{ij})\; .\\
\eeq
(e)
Comparing the integrand of the $\omega_{ij}$ integral 
with formula \eqref{regunitarity}, we identify it as the discontinuity in
$\omega_{ij}$ of the partial wave of the BFKL pomeron across the two-reggeon 
cut:
\bea
\label{discP}
\int \frac{d^2\bk_{i}}{(2 \pi)^2}
\gamma_2(\bk_{i},\bk_{j})
\phi^A(\bk_{i},\bq_{ij}-\bk_{i},\o_{ij})
\delta(\o_{ij}-\beta_i - \beta_j)
\phi^B(\bk_{i},\bq_{ij}-\bk_i;\o_{ij})= \nonumber \\
\frac{1}{\pi i} \disc_{\o_{ij}} [\cF_{BFKL}(\o_{ij},\bq_{ij})]
\eea
where 
\beq
\cF_{BFKL}(\o_{ij},\bq_{ij}) =
\int \frac{d^2 \bk d^2 \bk'}{(2 \pi)^4} \phi_0^A(\bk, \bq_{ij}-\bk) 
\cG_2(\bk,\bq_{ij}-\bk,\bk',\bq_{ij}-\bk';\omega_{ij}) 
\phi_0^B(\bk', \bq_{ij}-\bk').     
\eeq
Therefore \eqref{discFnb} becomes
\bea
\label{discFc}
\disc_\o \tF^n_k(\o,t) =
4 i \frac{(-1)^{n-k}}{k!(n-k)!}
\int d^2\bb e^{i \bb \cdot \bq} \nonumber \\
\cdot \prod_{(ij)}\Bigg[
\frac{1}{\pi i}
\int \frac{d^2\bq_{ij}}{(2 \pi)^2} ~ e^{-i \bb \cdot \bq_{ij}}
\int  d\o_{ij}
\disc_{\o_{ij}} [\cF_{BFKL}(\o_{ij},\bq_{ij})]\Bigg]
\delta(\o - \Sigma_{(ij)} \o_{ij}).
\eea
(f) Returning to the energy representation:
\bea
\tA_k^n(s,t) =
4 i s \frac{(-1)^{n-k}}{k!(n-k)!}\!
\int \!d^2\bb e^{i \bb \cdot \bq}
\Bigg[
\frac{1}{\pi i}
\int \frac{d^2\bq'}{(2 \pi)^2} ~ e^{-i \bb \cdot \bq'}
\!\int  \!d\o'
\disc_{\o'} [\cF_{BFKL}(\o',\bq')] s^{\o'}
\Bigg]^n ,
\eea
and defining:
\beq
\label{Omega}
\Omega(s,\bb)=
\frac{1}{\pi i}
\int \frac{d^2\bq}{(2 \pi)^2} ~ e^{-i \bb \cdot \bq}
\int d\o~
\disc_{\o} [\tF^n_{BFKL}(\o,\bq)] s^{\o},
\eeq
we obtain, after summation over $n \ge k$,
\beq
\label{Ak}
\tA_k(s,t) =
4 i s \int d^2\bb e^{i \bb \cdot \bq} P(s, \bb),
\eeq
where
\beq
P(s, \bb) = \frac{[\Omega(s,\bb)]^k}{k!}
e^{-\Omega(s,\bb)}
\eeq
is the probability of having $k$ cut Pomerons at fixed impact parameter $b$.\\ 
%%%%%%%%%%%%%%%%%%%%%%%%%%%%%%%%%%%%%%%%%%%%%%%%%%%%%%%%%%%%%%%%%%%%%%%%%%

%%%%%%%%%%%%%%%%%%%%% n-jets inclusive C.S. %%%%%%%%%%%%%%%%%%%%%%
\section{Single and double inclusive cross sections}
In this section we turn to the realistic case, where one ore more 
hard final states (e.g. jets or heavy flavors) are produced. Again we begin 
with a brief repeat of the AGK results and then turn to pQCD.    

In the AGK paper it has been shown that, for the single (or double) particle 
inclusive cross section, large classes of multi-pomeron corrections cancel. 
%figure------------------- 1 jet ------------------------
\begin{figure}[ht]
\begin{center}
\includegraphics[height=4cm]{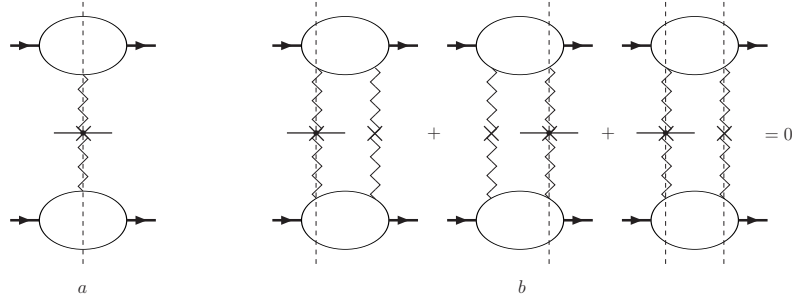}
\caption{The pattern of cancellation is shown for the single jet inclusive
case: the first graph survives, the three other cancel each other.}
\label{fig:1jetAGK}
\end{center}
\end{figure}
%figure--------------------------------------------------
We illustrate this result in Fig.\ref{fig:1jetAGK}:
for the single inclusive case (Fig.\ref{fig:1jetAGK}a) 
all multi-Pomeron exchanges across the produced particle cancel
(Fig.\ref{fig:1jetAGK}b), 
and the same is true for the double inclusive case
(Fig.\ref{fig:2jetAGK}a, \ref{fig:2jetAGK}b). 
For the latter case, however, there is a new  contribution: 
the produced particles originate from different Pomerons
(Fig.\ref{fig:2jetAGK}c). 
%figure------------------- 1 jet ------------------------
\begin{figure}[ht]
\begin{center}
\includegraphics[height=4cm]{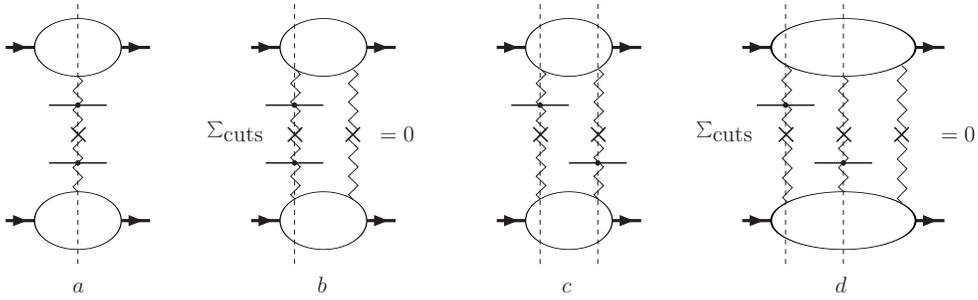}
\caption{The pattern of cancellation is shown for the double jet inclusive
case: the graph a and c contribute, the other interfere destructively
giving a vanishing contribution.}
\label{fig:2jetAGK}
\end{center}
\end{figure}
%figure--------------------------------------------------
This term is of particular 
interest, since it introduces longe range correlations in the 
rapidity difference $y_1 - y_2$: 
\beq
\rho(y_1, y_2) \sim  \frac{1}{\sigma_{tot}}\frac{d^2 \sigma}{dy_1
  dy_2} -
\frac{1}{\sigma_{tot}^2}
\frac{d\sigma}{dy_1} \frac{d\sigma}{dy_2} 
\eeq  
where, for simplicity, we have suppressed all variables other than the 
rapidities.
The multi-Pomeron corrections to this term, again, cancel
(Fig.\ref{fig:2jetAGK}d). 
In all cases, however, there remain multipomeron corrections between 
the production vertices and the projectiles. Examples are 
illustrated in Fig.\ref{surviving-terms}.  
%figure------------------- 1 jet ------------------------
\begin{figure}[ht]
\begin{center}
\includegraphics[height=4cm]{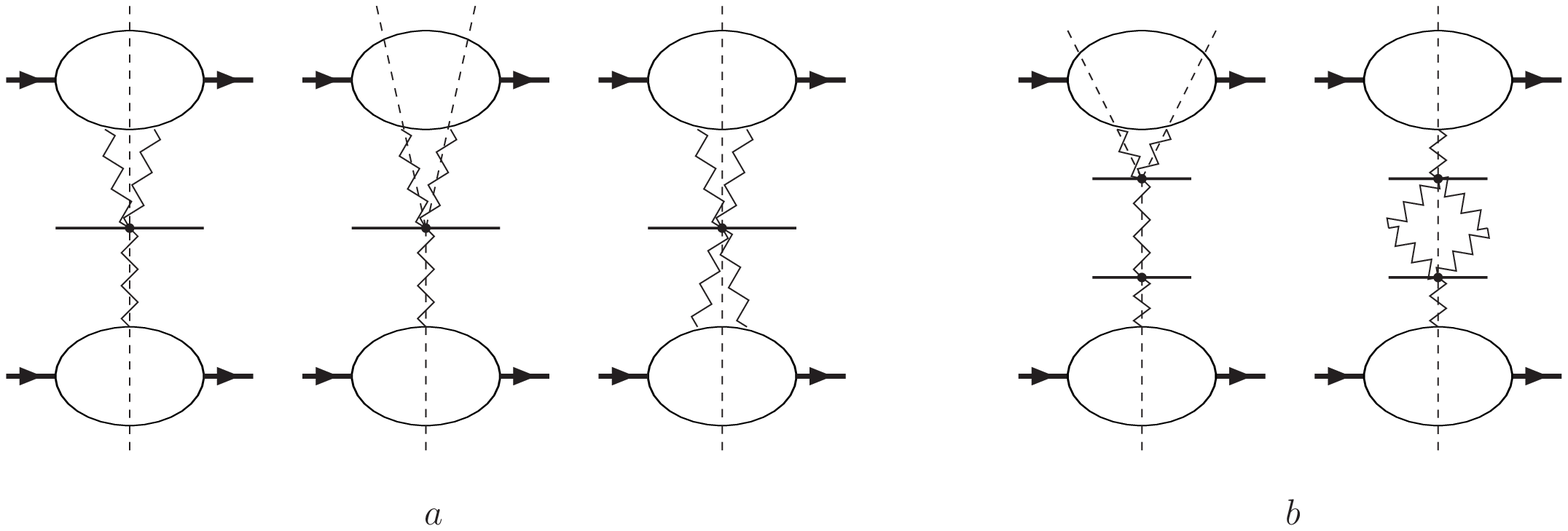}
\caption{Some examples of the terms that survive to the AGK cancellation for
the single jet production a) and for the double jet production b).}
\label{surviving-terms}
\end{center}
\end{figure}
%figure--------------------------------------------------

\subsection{Inclusive single jet production in pQCD}
We now turn to the corresponding final states in pQCD. 
As usual, the presence of the hard scale in the final state 
justifies, as far as the cut ladder with the produced jet or 
heavy flavor state is concerned, the use of perturbation theory.
%figure------------------- 1 jet ------------------------
\begin{figure}[ht]
\begin{center}
\includegraphics[height=4cm]{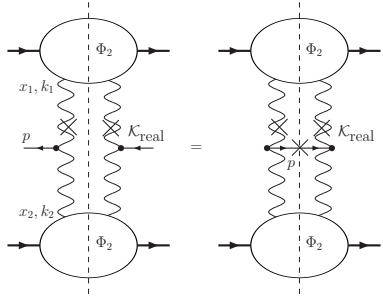}
\caption{Pictorial representation of the basic process for the single
inclusive jet production.}
\label{fig:1jetpQCD}
\end{center}
\end{figure}
%figure--------------------------------------------------
The basic process is illustrated in Fig.\ref{fig:1jetpQCD}: in the context of 
LL BFKL ladders (or $k_t$ factorization), 
the cross section has the form:  
\bea
\frac{d\sigma}{dy d^2\bp}(y,\bp) =
\frac{1}{4}
\int \frac{d^2\bk_1}{(2 \pi)^2}
\int \frac{d^2\bk_2}{(2 \pi)^2}
(2 \pi)^2\delta^{(2)}(\bp-\bk_1-\bk_2) \nonumber \\
\frac{\Phi_2(x_1; \bk_1, -\bk_1)}{\bk_1^2 \bk_1^2}
\cK_{\textrm{real}}(\bk_1,-\bk_1;-\bk_2,\bk_2)
\frac{\Phi_2(x_2; \bk_2, -\bk_2)}{\bk_2^2 \bk_2^2}
\label{1jetfirst}
\eea
with $x_1$ and $x_2$ being the momentum fractions 
of the incoming gluons with momenta $\bk_1$ and $\bk_2$, the rapidity
of the emitted jet given by $y=1/2 \log (x_1/x_2)$, and $\cK_{\textrm{real}}$
is the real emission BFKL kernel (Lipatov vertex) in the forward direction,
\beq
\cK_{\textrm{real}}(\bk_1,-\bk_1;-\bk_2,\bk_2) =
\frac{4 \alpha_s N_c}{(2 \pi)^2} \frac{\bk_1^2 \bk_2^2}{(\bk_1+\bk_2)^2}.
\label{realkernel}
\eeq
The connection between the unintegrated gluon densities 
$\Phi_2(x,\bk,-\bk)$ and the usual gluon density $g(x,Q^2)$ is given by: 
\beq
x g(x,Q^2) = \int_{Q_0^2}^{Q^2} \frac{d^2\bk}{\bk^2}
\Bigg[ \frac{2}{(2 \pi)^4} \Phi_2(x,\bk,-\bk) \Bigg].
\label{unint}
\eeq
 In $k_t$ factorization, the kernel \eqref{realkernel}
builds up the jet production subprocess in the approximation where
two reggeized gluons with momenta $\bk_{1,2}$ merge into a single
gluon with momentum $\bk_1+\bk_2$ which subsequently
originates the observed jet:
\beq
\frac{d \hat{\sigma}_{gg}(x_1, \bk_1, x_2, \bk_2 , y, \bp)}{dyd^2\bp} =
\frac{1}{2 s x_1 x_2}
\cK_{\textrm{real}}(\bk_1,-\bk_1;-\bk_2,\bk_2)
(2 \pi)^4 \delta^{(4)}(p-k_1-k_2) = \nonumber
\eeq
\beq
=\frac{(2 \pi)^4}{s^2 x_1 x_2}
\cK_{\textrm{real}}(\bk_1,-\bk_1;-\bk_2,\bk_2)
\delta^{(2)}(\bp-\bk_1-\bk_2)
\delta \Big( x_1-\frac{E_\perp(\bp^2)}{\sqrt{s}} e^y \Big)
\delta \Big( x_2-\frac{E_\perp(\bp^2)}{\sqrt{s}} e^{-y} \Big),
\label{1jethard}
\eeq
where $E_\perp(\bp^2)=\sqrt{m^2+\bp^2}$ is the transverse energy
of the jet and $m$ its invariant mass (in our case of a single gluon 
we have $m=0$).
Using \eqref{1jethard}, eq. \eqref{1jetfirst} can be cast in the form
\bea
\frac{d\sigma}{dy d^2\bp}(y,\bp) =
\frac{s^2}{4(2\pi)^2}
\int \frac{d^2\bk_1}{(2 \pi)^2}
\int \frac{d^2\bk_2}{(2 \pi)^2}
\int dx_1 \int dx_2 \nonumber \\
\frac{x_1 \Phi_2(x_1; \bk_1, -\bk_1)}{\bk_1^2 \bk_1^2}
\frac{d\hat{\sigma}_{gg}}{dy d^2\bp}(x_1, \bk_1, x_2, \bk_2 , y, \bp)
\frac{x_2 \Phi_2(x_2; \bk_2, -\bk_2)}{\bk_2^2 \bk_2^2}
\label{eq1jet}
\eea
which, by means of \eqref{unint}, strongly resembles the analogous formula
emerging in collinear factorization.

\begin{figure}[ht]
\begin{center}
\includegraphics[width=12cm]{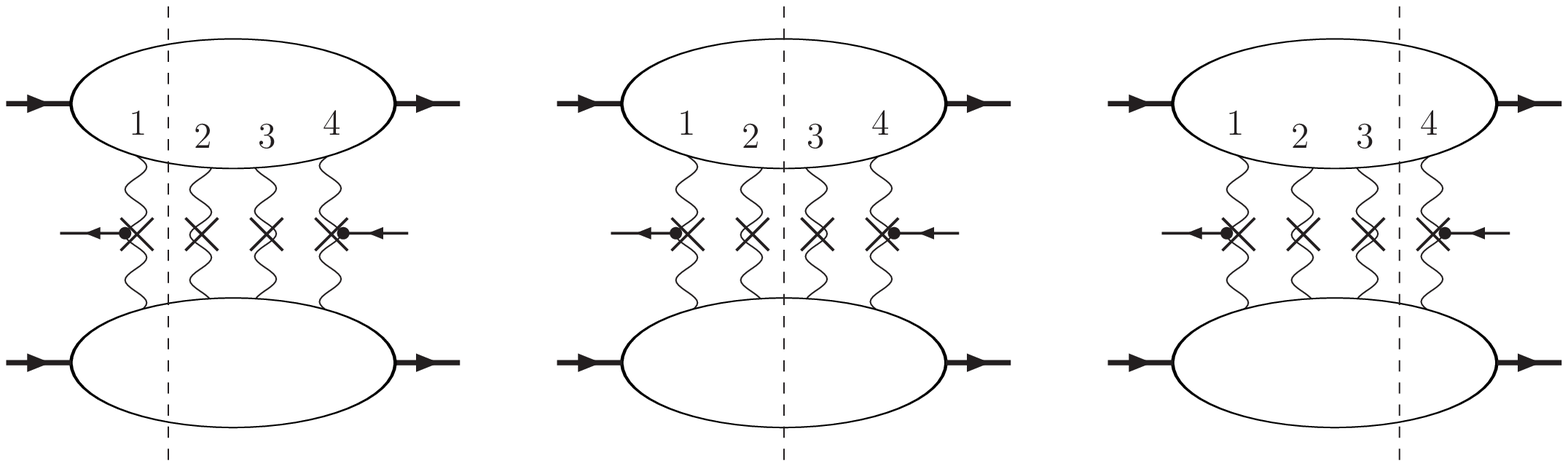}
\caption{Graphical representation of the process describing the
single inclusive cross section.}
\label{fig:1jet}
\end{center}
\end{figure}

\begin{figure}[ht]
\begin{center}
\includegraphics[width=15cm]{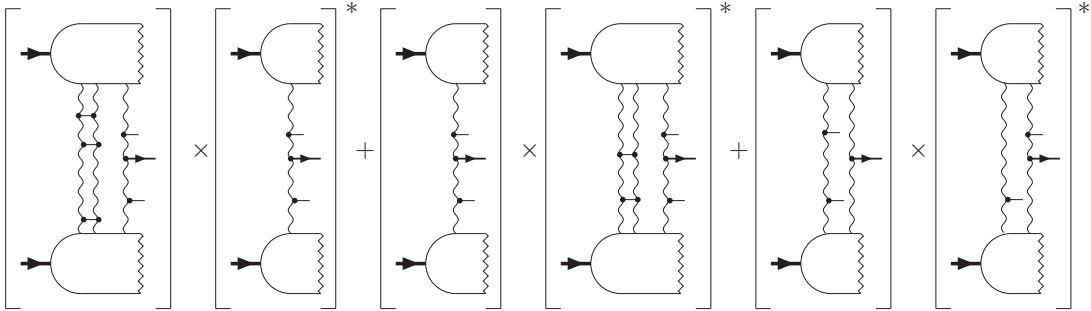}
\caption{An example of the interference among different processes that
produce the cancellation. The line with the arrow in the final state
correspond to the emission of the jet.}
\label{1jet-example}
\end{center}
\end{figure}

Next we consider the exchange of four reggeized gluons (Fig.\ref{fig:1jet});
the corresponding production processes are illustrated in
Fig.\ref{1jet-example}. The coupling of the gluons to the proton goes via
the functions $\cN_4$; compared with (\ref{eq1jet}) we restrict ourselves to 
the limit of small $x_1$, $x_2$. We do not write down formulae, 
but restrict ourselves to a qualitative discussion.  
%figure--------------------------------------------------
%figure--------------------------------------------------
%figure--------------------------------------------------
%figure--------------------------------------------------
Applying the counting arguments of the previous section to Fig.\ref{fig:1jet} 
we will show that all contributions sum up to zero. The symmetry factors
$1/2!$ etc. are the same as before; the counting of pairings is 
slightly different. In the previous case we have asked for 
multiplicities which has led us to count bound states above and below 
the $t$-channel intermediate state. Now we compute a one jet inclusive cross 
section and count the number of possibilities to attach the produced 
parton (jet). According to our assumptions, the couplings above and below 
are assumed to be symmetric under the exchange of gluon lines.   
In the first diagram of Fig.\ref{fig:1jet}, the number of ways of 
attaching the produced parton to the reggeons in the left is $1$, 
on the right $3$. Therefore, the symmetry factor $-1/3!$ 
must be multiplied by $3$. The same counting applies to the third diagram 
of Fig.\ref{fig:1jet}. For the diagram in the center we obtain $(1/2!)^2
\times 2^2$ which cancels against the other two diagrams. In summary, 
the four gluon corrections to the inclusive cross section
Fig.\ref{fig:1jetpQCD} cancel.

Several remarks have to made about this result. First, all arguments 
given above apply to the inclusive cross section: in Fig.16 we have 
illustrated, as an example, a few final states which contribute. 
These final states alone will not sum up to zero: the cancellations 
are valid only {\it after summation and integration over all final states
partons other than the parton singled out by the jet (marked by an arrow)}.
This means that, for individual events, these AGK cancellations are not 
visible. It is only after the summation over many events that the 
cancellations work. A necessary ingredient for this are the 
rescattering contributions (first and second term in Fig.\ref{1jet-example}):
if they would be left out, AGK would not work.
   
Second, the observed jet has introduced the hard scale which is necessary 
for justifying the use of gluon ladders. Such a hard scale is not present 
in the other ladders (e.g. in the uncut ladder in the first diagram of Fig.16, 
or in the cut ladder in the last graph of Fig.16): 
in the inclusive cross section we sum over final states which might include a 
large fraction of soft final states. So, strictly 
speaking, we have demonstrated only the cancellations between hard final 
states. This is certainly important for the modelling of multiple parton 
interactions which lead to the production of partons in the final states.
However, it is important to include also soft rescattering, 
i.e. the additional exchange of nonperturbative Pomerons. This can be done 
by combining the perturbative discussion of this subsection 
with the nonperturbative one given in subsection 3.1 There 
we have noted that, when adding soft Pomerons to a single cut 
Regge pole, the sum over all cuts across the soft Pomerons cancels.
If we simply substitute our hard cut Pomeron (containing the 
produced jet) for this single cut Regge pole,
we conclude that, in our pQCD inclusive cross section, 
also all additional soft Pomeron exchanges cancel. This coincides 
with the well known result in the collinear 
factorization which follows from the QCD factorization theorems~\cite{CSS}.

A final comment applies to multiple exchanges between the produced jet 
and one of the hadron projectiles. If one of the two momentum fractions, 
say $x_1$,  becomes very small, saturation effects are expected to  
become important: first corrections of this type are shown in
Fig.\ref{surviving-terms}.
Obviously, they require higher order jet production vertices which
have not yet been calculated. These vertices 
are somewhat analogous to the lowest order coupling of two BFKL ladders to the 
photon, i.e. to the process: photon + BFKL $\to$ quark-antiquark. Namely, 
if we open, in the first figure of Fig.\ref{surviving-terms}, 
the cut Pomeron below the jet vertex, we can view this vertex 
as the square of the subprocess: gluon + BFKL $\to$ jet. An important  
difference between the two cases lies in the fact that 
the incoming virtual photon is replaced by a colored gluon with transverse 
momentum $\bk_2$. In this context it is to be expected that the reggeization
of the gluon will be an issue: similar to our remarks on $\cD_4$ (second part 
of section 2, eq.(\ref{D4decomp})), there will be pieces which belong to 
antisymmetric two-gluon states and require a separate discussion.    
Another question of particular
interest is the applicability of the dipole picture: in contrast to the 
color singlet photon, the incoming gluon carries color and might lead to 
changes of the impact factor.

We conclude this section with the generalization of our discussion to an 
arbitrary number of reggeized gluons: in the appendix we show that 
the cancellation works for a general (even) number of additional 
reggeized gluons. This leads to the remarkable conclusion that 
there are no multi-Pomeron corrections to the basic process illustrated 
in Fig.14: all soft or hard exchanges between the upper and lower 
projectiles cancel. What remains are only the multiple exchanges between 
the produced gluon and the upper (or lower) proton.

\subsection{Inclusive double jet production}\label{2jets}
We now discuss the inclusive production of two jets, the Mueller-Navelet
cross section.
%figure--------------------------------------------------
\begin{figure}[ht]
\begin{center}
\includegraphics[height=5cm]{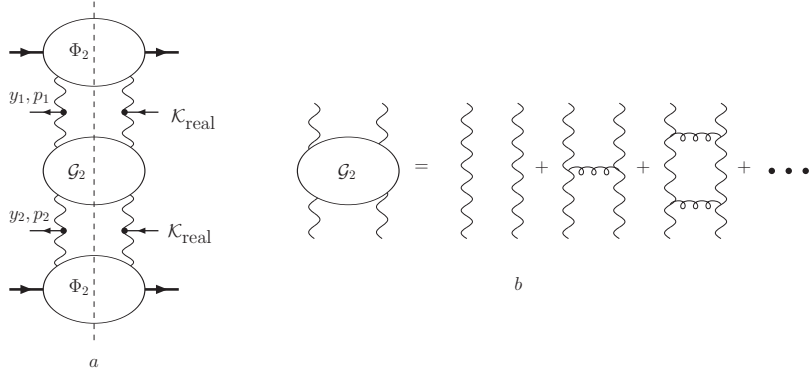}
\caption{a) Leading term for the inclusive production of two jets with a
large rapidity interval between them. b) Definition of the two-to-two
reggeized gluons BFKL Green function.}
\label{2jets-2reggeons}
\end{center}
\end{figure}
%figure--------------------------------------------------
The leading contribution is shown in Fig.\ref{2jets-2reggeons}a: when the
rapidity interval between the two observed jets is large the process
is described by Regge kinematics, and the large rapidity interval is due
to the BFKL Green function $\cG_2$, which contains the exchanges of
gluons between the reggeized gluons (Fig. \ref{2jets-2reggeons}b);
note that, in our convention, $\cG_2$ does not contain the propagators
for the external gluons.

The formula for the cross section associated with Fig.\ref{2jets-2reggeons}a is
\beq
\frac{x_1 x_2 d^2\sigma}{dx_1 dx_2 d^2\bp_1 d^2\bp_2}
(x_1,x_2,y,\bp_1,\bp_2) =
\frac{1}{4}
\int \frac{d^2\bk}{(2 \pi)^2}~ \frac{d^2\bk'}{(2 \pi)^2} \nonumber
\eeq
\beq
\frac{\Phi_2( x_1; \bk, -\bk )}{\bk^2\bk^2}
\cK_{\textrm{real}}(\bk,-\bk;\bk-\bp_1,-\bk+\bp_1) \nonumber
\eeq
\beq
\frac{{\cal G}_2
( y ; \bk-\bp_1 , -\bk+\bp_1 ; -\bk'+\bp_2 , \bk'-\bp_2 )}
{(\bk-\bp_1)^2 (\bk-\bp_1)^2 (\bk'-\bp_2)^2 (\bk'-\bp_2)^2}
\label{MNnf}
\eeq
\beq
\cK_{\textrm{real}}(\bk',-\bk';\bk'-\bp_2,-\bk'+\bp_2)
\frac{\Phi_2( x_2; \bk', -\bk' )}{\bk'^2\bk'^2}. \nonumber
\eeq
Defining $\tilde{\cal G}_2$ as the usual Green function but with the
propagators for the gluons on the left side of the cut,
\beq
\tilde{\cal G}_2(y; \bk_1,\bk_2; \bk'_1,\bk'_2 ) =
\frac{{\cal G}_2(y; \bk_1,\bk_2; \bk'_1,\bk'_2 )}{\bk_1^2 \bk'^2_1}
\label{green_scaled}
\eeq
and using \eqref{realkernel}, \eqref{MNnf} can be rewritten as
\beq
\frac{x_1 x_2 d^2\sigma}{dx_1 dx_2 d^2\bp_1 d^2\bp_2}
(x_1,x_2,y,\bp_1,\bp_2) = \nonumber
\eeq
\beq
\frac{1}{4 (2 \pi)^4}
\frac{1}{\bp_1^2 \bp_2^2}\bigg[\frac{4 \alpha_s N_c}{(2 \pi)^2}\bigg]^2
\int d^2\bk
\frac{\Phi_2( x_1; \bk, -\bk )}{\bk^2}
\int d^2\bk'
\frac{\Phi_2( x_2; \bk', -\bk' )}{\bk'^2}
\label{MNnfb}
\eeq
\beq
\tilde{\cal G}_2
( y ; \bk-\bp_1 , -\bk+\bp_1 ; -\bk'+\bp_2 , \bk'-\bp_2 ). \nonumber
\eeq
Note that in practice, because of the limited energy of the hadron 
collider (e.g. the Tevatron), the kinematics of the Mueller-Navelet is 
chosen such that it maximizes the rapidity gap 
between the two jets; this implies that the momentum fractions $x_1$ and 
$x_2$ are not necessarily small, i.e. the separations in rapidity between 
the jets and the projectiles $A$ and $B$ are not particularly large.
As a result, we have strong $k_t$ ordering between the jets and the 
projectiles, and we are led to the usual integrated gluon densities
(eq.(\ref{unint})). The cross section \eqref{MNnf} can be cast in the form
\beq
\frac{x_1 x_2 d^2\sigma}{dx_1 dx_2 d^2\bp_1 d^2\bp_2}
(y,x_1,x_2,\bp_1,\bp_2) = \nonumber
\eeq
\beq
=
x_1 g(x_1,M^2)~
\bigg[\frac{\alpha_s N_c}{\bp_1^2}\bigg]
\tilde{\cal G}_2(y;-\bp_1,\bp_1;\bp_2,-\bp_2)
\bigg[\frac{\alpha_s N_c}{\bp_2^2}\bigg]~
x_2 g(x_2,M^2)
\label{eq:MN}
\eeq
which is the well known Mueller-Navelet formula for dijet production in
hadron-hadron scattering \cite{MN}. Also, the incoming gluons with momenta 
$\bk_1$ and $\bk_2$ may be replaced by quark lines, and we have additional  
contributions from the quark densities. Throughout our discussion of the 
AGK counting, however, we restrict ourselves to the region of small 
$x_1$ and $x_2$. 

The first correction to \eqref{eq:MN} comes from the exchange of four
reggeized gluons. Our discussion requires small $x_1$ and $x_2$, and the
coupling of the four gluons proceeds via the function $\cN_4$. 
The situation is akin to that shown in Fig.\ref{fig:1jet},
but now there are two vertices for the jet emission on each side of the cut.
Between the two vertices, the four reggeized gluons interact exchanging gluons;
this interaction is symmetric under the exchange of two gluon lines. 
We denote this interaction by the Green's function ${\cal G}_4$ (which
technically is obtained from the corresponding BKP
kernel \cite{BKP,Jaroszewicz} describing the evolution of $n$ interacting
reggeized gluon states in the $t$-channel).
Following the same line of counting as before, we will find that only a 
particular subset of the terms gives a non vanishing contribution to the 
inclusive observable.
%figure--------------------------------------------------
\begin{figure}[ht]
\begin{center}
\includegraphics[width=10cm]{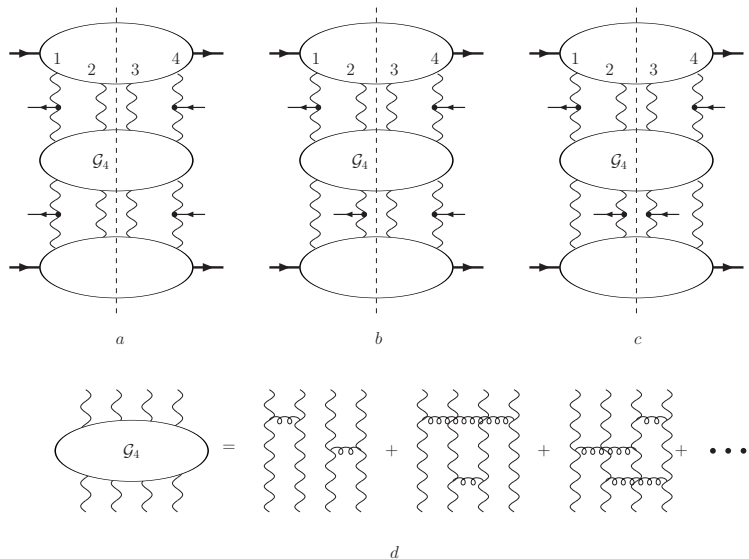}
\caption{We show the central cut as an example of the classification
of the different terms which a priori should contribute to the cross section:
a) both vertices attached to the same gluon on both side, b) to the same
gluon on one side but to different gluons to the other, and c) to different
gluons on both sides. In d) we show the definition of the four-to-four
reggeized gluons Green function.}
\label{fig:2jets}
\end{center}
\end{figure}
%figure--------------------------------------------------
We treat  three different cases, which are illustrated in 
Fig.\ref{fig:2jets} (where only the central cut case is shown):\\
(a) Both production vertices are attached to the same reggeized gluons 
on each side of the cut; in this case the counting is exactly the same as 
in the previous section, and the sum of the correction terms vanishes.\\
(b) Both vertices are attached to the same reggeized gluon on one side
of the cut, but to different reggeized gluons on the other side.
For example, suppose that the vertices
are connected to the same reggeized gluon on the right side
(the other case is identical).
The first cut (between $1$ and $2$) does not contribute since there is only
one reggeized gluon on the left side. 
The combinatorial factor for the central cut (between $2$ and $3$) is $2$
(the reggeized gluon whom to attach the right sides of the vertices in the
right side of the cut can be chosen among two) times $2$
(the left side of the first vertex can be attached to one of the two
reggeized gluon on the left, but the other vertex can only be attached
to the one left); together with
the symmetry factor $1/4$, the contribution of the diagram is $1$.
Finally, the combinatorial factor of the third
cut (between $3$ and $4$) is $3 \times 2$ because one of the vertices
can be attached to one of the three reggeized gluons on the left but
the other one to one of the two not connected to the first one.
Combining this with the symmetry factor $-1/6$, the contribution
of the diagram becomes $-1$,
which cancels exactly the $+1$ of the central cut.\\
(c) The vertices are attached to different reggeized gluons on both sides of
the cut: only one diagram contributes, and there are no cancellations.

What emerges from this analysis of the four gluon exchange is that those 
diagrams, in which there are reggeized gluons without a production vertex,
give a vanishing contribution. This statement holds for an arbitrary number of 
reggeized gluons and any number of jets; in particular, diagrams
involving the exchange of more that four reggeized gluons do not contribute
to the double inclusive jet production. A general proof is given in the 
Appendix. 

%figure--------------------------------------------------
\begin{figure}[ht]
\begin{center}
\includegraphics[height=5cm]{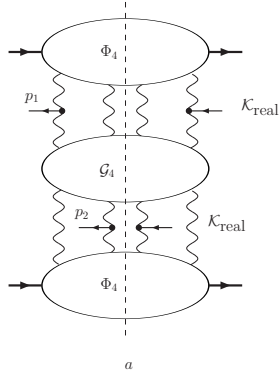}
\caption{a) Non vanishing correction to the two jets inclusive jet production
due to the exchange of four reggeized gluons.}
\label{2jets-4reggeons}
\end{center}
\end{figure}
%figure--------------------------------------------------
For the case of the Mueller-Navelet jets we are thus left with only one 
type of corrections (Fig.\ref{2jets-4reggeons}): the jets are emitted
from different reggeized gluons on both sides of the cut.
We write the explicit formula for the correction represented in
Fig.\ref{2jets-4reggeons} in the following form:
\beq
\frac{x_1 x_2 d^2\sigma}{dx_1 dx_2 d^2\bp_1 d^2\bp_2}(y,x_1,x_2,\bp_1,\bp_2) =
\frac{1}{4}
\int \Bigg[ \prod_{i=1}^4
\frac{d^2\bk_{i}}{(2 \pi)^2}
\frac{d^2\bk'_{i}}{(2 \pi)^2} \Bigg]
(2 \pi)^2\delta^{(2)}(\Sigma_i \bk_i)
(2 \pi)^2\delta^{(2)}(\Sigma_i \bk'_i) \nonumber
\eeq
\beq
\frac{\Phi_4(x_1;\bk_1,\bk_2,\bk_3,\bk_4)}
{\bk_1^2 \bk_2^2 \bk_3^2 \bk_4^2}~
\frac{\cK_{\textrm{real}}(\bk_1,\bk_4;\bk_1-\bp_1,\bk_4+\bp_1)}
{(\bk_1-\bp_1)^2 (\bk_4+\bp_1)^2} \nonumber
\eeq
\beq
{\cal G}_4 \bigg(y;
\begin{array}{c}
\bk_1-\bp_1 , \bk_2        , \bk_3        , \bk_4+\bp_1 \\
\bk'_1      , \bk'_2-\bp_2 , \bk'_3+\bp_2 , \bk'_4
\end{array}
\bigg)~
\label{MN4g}
\eeq
\beq
\frac{\cK_{\textrm{real}}(\bk'_2,\bk'_3;\bk'_2-\bp_2,\bk'_3+\bp_2)}
{(\bk'_2-\bp_2)^2 (\bk'_3+\bp_2)^2}~
\frac{\Phi_4(x_2;\bk'_1,\bk'_2,\bk'_3,\bk'_4)}
{\bk'^2_1 \bk'^2_2 \bk'^2_3 \bk'^2_4} \nonumber
\eeq
Here we have used another notation for the coupling of the reggeons 
to the proton, $\Phi_4$ rather than $\cN_4$: in the limit of small 
$x_i$,  $\Phi_4$ coincides with  $\cN_4$. In a more realistic 
situation we might allow for strong ordering of transverse momenta 
in those reggeized gluons which connect the jet vertex (cf.our discussion 
after \eqref{MNnfb}). As a consequence, the pattern of integrations 
in \eqref{MN4g} is more involved than it was in \eqref{MNnfb}.
In particular, the relation between the coupling $\Phi_4$ 
and the vertex functions $\cN_4$ requires a more detailed discussion.

We finally turn our attention to the Green function $\cG_4$; since
only the central cut survives, there are only diagrams corresponding
to multiplicity $k=0,2$. The diffractive case (i.e. the first diagram
drawn in Fig.\ref{fig:2jets}d) is known in the literature as ``hard color
singlet'' \cite{HCS}: there is a rapidity gap between the two observed jets.
Performing the usual counting it is trivially verified
that the weight factors of the double multiplicity part of
the cut Green function ($k=2$) is twice the one of the diffractive part
($k=0$) and, as before, a switch of the pairing structure 
inside $\cG_4$ does not spoil this relation.

Specific models ~\cite{Kovchegov1_2,Braun2005} require the discussion 
of reggeon diagrams which are more complicated than those discussed in this 
paper; in particular, this includes diagrams which change the number 
reggeized gluon in the $t$-channel. For any $t$-channel state with 
a fixed number of gluons, $2n$, our discussion applies. However, when 
considering, for example, double inclusive jet cross sections where 
between the two emissions the number of gluons changes, one expects to see 
new contributions where one of the produced jet originates from the 
gluon number changing vertex. As we have said above, the application of 
of AGK cutting rules to such vertices requires a separate discussion.          

\subsection{Remarks on the relation to phenomenological models}

The topic of multiple interactions has been addressed for many years
(see, for example, \cite{SS} and references therein, or \cite{RT}).
Key elements are multi-parton distributions, which are interpreted as 
probabilities of finding, inside the hadron, a number
$n_p \ge 1$ of partons with longitudinal momentum fractions 
$\{x_i\},i=1,...,n_p$. In the framework of $k_t$-factorization 
the partons also carry transverse momenta $\{k_i\}$. 
These partons then interact through hard subprocesses 
and produce partonic final states. In the final step, color strings between 
the produced partons and the remnants of the hadrons describe the 
hadronization.

The AGK analysis of this paper has mostly been formulated in terms of angular
momenta, which are conjugate to energy variables. When translating our 
results, one first has to emphasize that all the AGK analysis applies to 
the limit of small momentum fractions $x_i$. In particular, in the coupling 
functions $\cN_n$ the gluons have longitudinal momentum
fractions $x_i$ which are all small and defined to be all of the same order.
Hence, in our notation we have not specified the dependence upon the
$x_i$ but rather used the conjugate variable of total angular momentum 
$\omega$. 
However, as we have discussed after eq.(\ref{gammafactor}) and 
illustrated by our example of the eikonal model, 
it is possible to define more general coupling functions $\cN$ by introducing 
a dependence upon angular momenta of subsystems of gluons, e.g. $\omega_{ij}$. 
This corresponds to momentum fractions which are small but of different 
order of magnitude. But, as usual 
in the small-$x$ approximations, conservation of longitudinal momenta
is not observed, and therefore a detailed assignment of longitudinal 
momenta is not easy. Therefore, as far as the modelling of multiple 
interactions is concerned, 
the AGK analysis presented in this paper should mainly be 
viewed as providing constraints for the limit of small-$x$ values.
For a hadron collider which operates at energies as high as the LHC, 
these constraints should be quite essential.    
%figure--------------------------------------------------
\begin{figure}[ht]
\begin{center}
\includegraphics[height=5cm]{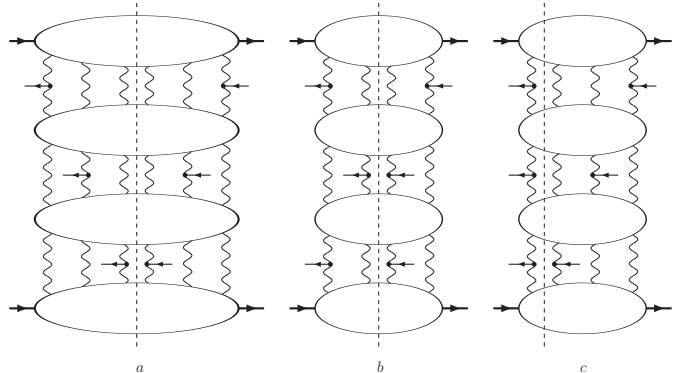}
\caption{Some diagrams contributing to the three jet production
  notwithstanding the AGK cancellations.}
\label{3jet}
\end{center}
\end{figure}
%figure--------------------------------------------------

From the AGK analysis it becomes quite clear that 
there exist contributions to the cross section which cannot 
simply be interpreted as probability distributions.  
Examples have been illustrated in Fig.16: whereas the
third contribution represents a square of production amplitudes, and 
the couplings of the gluons to the upper (or lower) proton might be 
interpreted as a 'probability of finding two gluons inside a proton',
the first and the second contributions are interference terms.
Another example of a probabilistic contribution is shown in 
Fig.\ref{2jets-4reggeons}. The third diagram in Fig.20, on the other hand, 
illustrates an another interference term which survives after all AGK 
cancellations have been worked out. 

One is therefore lead to a more general concept of {\it multiparton 
correlators}. Even in the framework of collinear
factorization at the leading twist level, there is already the well-known 
example of 
the \emph{Generalized Parton Distributions} (see \cite{GPD} for a review),
which represent correlations between two partons inside the proton.
In this case it is possible to obtain direct experimental information 
of these correlations by exploiting, for
example, the interference between the DVCS
(Virtual Compton Scattering) and the BH (\emph{Bethe-Heitler}) 
process \cite{DVCS}. Beyond these two-parton correlators 
very little has been worked out in the literature.  
%%%%%%%%%%%%%%%%%%%%%%%%% conclusions %%%%%%%%%%%%%%%%%%%%%%%%%%%%
\section*{Conclusions}
In this paper we have discussed the origin and a few consequences of the 
AGK cutting rules in pQCD where the Pomeron is described by the 
BFKL gluon ladders. We have identified the basic ingredient for the 
validity of the AGK cutting rules, the symmetry properties of 
couplings of $n$ reggeized gluons to the projectile; they are in 
agreement with general properties of reggeon unitarity equations,
and they have explicitly been verified in pQCD in $\gamma^* \gamma^*$ 
(or onium-onium scattering). 

As to consequences of the AGK counting rules, we have considered 
single and double inclusive jet production. In both cases,
multipomeron exchanges across the produced jets cancel; this holds 
for both soft and hard Pomerons. In the two-jet inclusive case, there 
exists an extra contribution in which the jets are emitted from two 
different parton chains. As an example, we have analysed how this 
contribution looks like in in the Mueller-Navelet jet final states;
this contribution turns out to be related also to the hard color singlet 
exchange cross section.

At very small $x$ values, QCD saturation effects are expected to become 
important. Our AGK analysis leads to the conclusion that such nonlinear 
corrections should be modelled as multiple exchanges between the jet vertices 
or between jet vertices and either of the projectiles, 
but not between the two projectiles.

Several important questions have not been addressed in this paper,
and our analysis of AGK counting rules in proton proton scattering 
remains incomplete.  
First, QCD allows two-gluon states in the $t$-channel that are 
antisymmetric under 
momentum and color exchange: in pQCD, bootstrap equations have been shown 
to be valid, as a result of which the such two-gluon states are identical 
to a single reggeized gluon. We conjecture that this property holds also 
beyond pQCD: in this case, as far as the reggeization of gluons is 
concerned, we do not need to introduce new couplings to the hadron, and our 
discussion can be reduced to the symmetric functions which ensure the validity 
of the AGK counting rules. Nevertheless, the existence of these 
reggeizing pieces require further studies: the most prominent example 
is $\gamma^*$-proton scattering (i.e. the deep inelastic structure 
function), where in leading order the coupling of four reggeized 
gluons to the virtual photon consist of reggeized pieces only. This 
process will be studied in a forthcoming paper.  
       
In this paper we also have not yet included the Odderon. Perturbative 
QCD contains an Odderon state, built from (at least) three interacting
reggeized gluons. It requires couplings to the proton which satisfy
special symmetry properties. 
Since the intercept of the highest Odderon state known so far 
~\cite{BLV} lies at unity, at high energies it will be less important;
a complete analysis, however, will have to include Odderon exchange.
  
Another aspect which requires further studies is the limitation 
due to finite energies. Strictly speaking, the AGK analysis in terms 
of ladder diagrams requires infinite energies: at any finite energy,
the number of produced gluons inside a cut ladder is limited, and, 
therefore, also the number of different 'cut ladders', $k$, cannot be 
arbitrarily large. When modelling multiple parton interactions, 
this may have consequences (e.g. incomplete cancellations) which have to 
worked out in detail.
%%%%%%%%%%%%%%%%%%%%%%%%%%%%%%%%%%%%%%%%%%%%%%%%%%%%%%%%%%%%%%%%%%
\section*{Acknowledgements}
M.S. is grateful for the kind hospitality to the \emph{II Institut
f\"ur Theoretische Physik} of the Hamburg University where part
of this work was done and the \emph{Programma Marco Polo} for
the financial support. G.P.V. is very grateful to the Hamburg
University for the warm hospitality.
We are grateful to Markus Diehl for his numerous helpful comments.
%%%%%%%%%%%%%%%%%%%%%%%%% Appendix %%%%%%%%%%%%%%%%%%%%%%%%%%%%%
\newpage
\section*{Appendix}
In this appendix we present a few details of the 
generalization of our results to $2n$ gluons.
We start with a few notations.
Let $\tA^n(s,t)$ denote the energy discontinuity of the contribution 
of $n$-reggeized gluon exchange in the elastic amplitude for the process 
$AB \to AB$, $\tF^n(\o,\bq)$.The energy discontinuity can be written as the 
sum over 
products of production amplitudes $\tilde{\cT}_{AB \to X}^{j}(s,t)$ 
for the process $AB \to X$, where the index $j$ stands for the number 
of exchanged reggeized gluons. Each such term is of the form 
$\tilde{\cT}_{AB \to X}^{j}(s,t) (\tilde{\cT}_{AB \to X}^{2n-j}(s,t))^*$, 
i.e. $j$ gluons are exchanged inside the ``left''
amplitude, $2n-j$ in the ``right'' complex conjugate amplitude,
and the total number of exchanged gluons is $2n$. 
In order to construct $\tA^n(s,t)$, 
we have to sum over $j=1,...,2n-1$.
Furthermore, we also have to sum (denoted by 
$\sum_\textrm{B.S.}$) over all the possibilities to form bound states 
of two-reggeized gluons bound states (pomerons) inside the amplitudes 
$\cN_{2n}$:
\beq
\tA^n(s,t) \stackrel{\textrm{def}}{=}
2 i \sum_{\textrm{B.S.}}\sum_{j=1}^{2n-1}
\int d \Omega_X~
\tilde{\cT}_{AB \to X}^{j}(s,t_1)~
\big(\tilde{\cT}_{AB \to X}^{2n-j}(s,t_2)\big)^*
= \nonumber
\eeq
\beq
=2 i \sum_{\textrm{B.S.}}\sum_{j=1}^{2n-1}
\int d\Omega_{2n}~s^{1+\tilde{\beta}}~ \cN_{2n}~ \cN_{2n}~
\frac{i\xi_\mathds{G}(\bk_1)...i\xi_\mathds{G}(\bk_j)}
{j!}
\frac{[i\xi_\mathds{G}(\bk_{j+1})...i\xi_\mathds{G}(\bk_{2n})]^*}
{(2n-j)!}
\label{discspQCD}
\eeq
Here we have made used of the fact that $\cN_{2n}$ does not depend on the 
position of the cut. The tilde symbol indicates that we are working in pQCD. 
Since the signature
factor $\xi_\mathds{G}(\bk)= - 2/\pi \bk^2$ is real
(cf. the discussion after eq.(13)),
\eqref{discspQCD} can be cast into the form
\beq
\tA^n(s,t) = 2^n \pi i \sum_{\textrm{B.S.}} \sum_{j=1}^{2n-1} S^n_j~
\int d\Omega_{2n}~s^{1+\tilde{\beta}_n}~ \cN_{2n}~ \cN_{2n}~
\tilde{\gamma}_{\{\bk_l\}}
\label{discspQCD1}
\eeq
where we have used \eqref{pQCDgamma}, and we have introduced the symmetry 
factor
\beq
S^n_j=\frac{(-1)^{n-j}}{j!(2n-j)!}.
\label{symfac}
\eeq

\subsection*{Inclusive case}
We begin with the fully inclusive case of section 3.
Starting from \eqref{discspQCD} we want to find a decomposition in terms 
of cut Pomerons.
For each position of the cut (denoted by $j$) we first classify all the 
different possibilities of forming $2$-gluon bound states out of the $2n$
reggeized gluons; at the same time we have to keep track of
how many of these bound states are cut by the energy cutting line.
Let $k$ denote the number of such cut bound states: then $k$ also 
labels the multiplicity of $s$-channel gluons intersected by the 
cutting lines. We then must count how many different
configurations $(i_1j_1)...(i_nj_n)$ contribute to a term belonging to  
the multiplicity $k$.

Denoting by $C^n$ the number of different pairings among $2n$ reggeized 
gluons, is clear that $C^n=(2n-1)!!$:
starting with the first reggeized gluon to the left of the cutting line, 
its partner can be chosen among $2n-1$ other $t$-channel gluons; 
the partner of the next unpaired reggeized gluon can be chosen among $2n-3$ 
other gluons, and so forth until all the reggeized gluons have been put into 
pairs. Next we will decompose $C^n$ into contributions with 
fixed-multiplicity $k$:
\beq
C^n = \sum_k C^n_{jk},
\label{cnsum}
\eeq
where $C^n_{jk}$ is the number of configurations that
contain exactly $k$ cut pairs. The subscript $j$ indicates that this 
number will depend upon the position of the cutting line.
It is clear that $C^n_{jk}$ is $0$ unless $k$ and $j$ are both even or both 
odd, i.e. $C^n_{jk} \propto 1/2(1+(-1)^{j+k})$. Also, we need $k\le j$ and 
$k\le 2n-j$, i.e. the range of $k$ values in \ref{cnsum} depends upon $j$.
If $j$ and $k$ satisfy these conditions
we can chose $k$ reggeized gluons in each side of the cut in $\binom{j}{k}
\binom{2n-j}{k}$ different ways and couple each reggeized gluon in the left 
side with a reggeized gluon in the right side (this can be done in $k!$ ways).
We are left with $j-k$ reggeized gluons on the left side and $2n-j-k$
on the right side, which must be coupled in pairs without crossing the cut; there are
$(j-k-1)!!(2n-j-k-1)!!$ ways to do that. Eventually the expression obtained is
\beq
C^n_{jk} = \frac{1+(-1)^{j+k}}{2}\frac{j! (2n-j)!}{k!(j-k)!!(2n-j-k)!!},
\label{cmjk}
\eeq
and it is easy to verify that \eqref{cmjk} satisfies \eqref{cnsum}.

With this expression for $C^n_{jk}$ we can rewrite 
\eqref{discspQCD} as a sum over the multiplicity index $k$
of fixed-multipicity contributions $\tA_k^n(s,t)$:
\beq
\tA^n(s,t) = \sum_{k=0}^n \tA_k^n(s,t)
\label{pQCDAGKdec}
\eeq
where
\beq
\tA_k^n(s,t) = 2^n \pi i \sum_{j=k}^{2n-k} S^n_j C^n_{jk}~
\int d\Omega_{2n}~s^{1+\tilde{\beta}}~
\cN_{2n}~ \cN_{2n}~
\tilde{\gamma}_{\{\bk_l\}}
\label{pQCDAk}
\eeq
The reader should note that the integral in \eqref{pQCDAk} does
not depend upon $j$ or $k$; therefore there is just a combinatorial factor
in front of the integral. Performing the sum in \eqref{pQCDAk} we arrive 
at:
\bea
\sum_{j=k}^{2n-k} S^n_j C^n_{jk} &=&
\frac{(-1)^n}{2~k!}\sum_{j=k}^{2n-k}
\frac{(-1)^j+(-1)^k}{(j-k)!!(2n-j-k)!!} \nonumber \\
&=& \Bigg\{ \begin{array}{ll}
\frac{(-1)^n}{n!} ( 1-2^{1-n} ) & \textrm{if $k=0$} \\
\frac{(-1)^{n-k}}{k!(n-k)!}     & \textrm{if $k>0$}
\end{array}.
\label{cutcounting}
\eea
With the definition \eqref{tildeF}, we have shown that
\eqref{pQCDAk} can be written as in \eqref{agkpQCD}.

\subsection*{Switches of Pairings}
Now we want to show that a switch from a configuration
$(i_1j_1) ... (i_nj_n)$ to $(i'_1j'_1) ... (i'_nj'_n)$
preserves the relative weight between different multiplicities $k$.
The situation is the following: at some point in rapidity we have the 
configuration $(i_1j_1) \dots (i_nj_n)$ with muliplicity 
$k$, and the weight factor of this multiplicity $k$ is the usual $\tF^n_k$
given in \eqref{tildeF}; moving now to a neighbouring rapidity interval   
we switch to a different configuration $(i'_1j'_1) \dots (i'_nj'_n)$,
and we want to compute the weight of the term with $k'$ cut rungs.
Contributions to this $k'$ can come from different $k$ terms before the
switch; we therefore must sum over $k$ the $\tF^n_k$ muliplied by the 
number of ways of obtaining a configuration with $k'$ cut rungs 
(normalized by the total number).

Put in other words, we must build an $(n+1) \times (n+1)$ matrix, where the 
initial multiplicity $k$ labels the columns, and the final multiplicity 
$k'$ labels the rows; the elements ${\cal M}^n_{k'k}$ of this
matrix are defined to be the fraction of configurations 
which, after the switch, lead to
$k'$ cut rungs. We then want to show that the vector $\tF^n_k$ is
an eigenvector of this matrix ${\cal M}^n_{k'k}$.
Using a vector notation and dropping the index $n$ we want to show that:
\beq
\mbf{\cF} \propto {\cal M} \mbf{\cF}.
\label{eq:ev}
\eeq
But $\mbf{\cF}$ is an eigenvector of ${\cal M}$ iff it is an eigenvector
of ${\cal M}+c \mathds{1}$ where $c$ is an arbitrary constant.
Note that, by definition, ${\cal M}$ is computed by considering only 
transitions between \emph{different pairwise configurations}, and a 
transition to the same configuration, in our matrix notation, 
is proportional to the identity (the nature of the cut does not change).
The proportionality constant is just $1/C^n$, where $C^n$ is the total number
of configurations defined before.
We normalize the matrix ${\cal M}$ by dividing by $C^n-1$
instead of $C^n$, because
we are considering only transitions to different configurations; of course,
this normalization does not affect the form of the eigenvectors (it changes
only the eigenvalues). So we can change its normalization multiplying by
$(C^n-1)/C^n$. Eventually, proving \eqref{eq:ev} is equivalent to prove
\beq
\mbf{\cF} \propto \overline {\cal M} \mbf{\cF},
\label{eq:ev2}
\eeq
where we have introduced the new matrix $\overline {\cal M}$ defined by
\beq
\overline {\cal M} = \frac{(C^n-1) {\cal M} + \mathds{1}}{C^n}.
\eeq
This new matrix $\overline {\cal M}$ now contains \emph{all possible switches}
to a new configuration, not only to the different ones.

Instead of computing explicitely the coefficients of $\overline {\cal M}$,
it is easier to compute directly the RHS of \eqref{eq:ev2}, i.e.
write the sum over the various contributions due to different positions of 
the cut before the switch. For fixed $j$ and $k$ it is trivial to obtain 
the fraction of configurations with $k'$ cut rungs: it is $C^n_{jk'}/C^n$ 
(note that this is true only because we are no longer restricting ourselves
to the new configuration being different from
the previous one). Therefore, using
\beq
\tF^n_k = 2^{n-1} n!~
\sum_{j=k}^{2n-k} S^n_j C^n_{jk},
\eeq
the rhs of \eqref{eq:ev2} can be written as
\beq
\sum_{k=0}^{n}
2^{n-1} n!~
\sum_{j=k}^{2n-k} \frac{C^n_{jk'}}{C^n} S^n_j C^n_{jk}.
\eeq
After using \eqref{cnsum}, this gives exactly $\tF^n_{k'}$, i.e. 
the $\tF^n_{k}$ form an eigenvector.

\subsection*{$1$-Jet inclusive}
In this subsection we generalize the cancellation of diagrams with
$n>2$ gluon lines in the single inclusive cross section. The discontinuity 
of the amplitude for single jet production is
given by
\beq
\tA^n_{1\textrm{-jet}}(s,t) =
2 i \sum_{\textrm{B.S.}}\sum_{j=1}^{2n-1} j(2n-j) S^n_j~
\int d\Omega_{2n}~s^{1+\tilde{\beta}}~ \cN_{2n}~ \cN_{2n}~
\tilde{\gamma}_{\{\bk_l\}},
\label{discspQCD1jet}
\eeq
where the factor $j(2n-j)$ counts the number of ways in which the jet can be
connected to the reggeized gluons on the lhs and rhs of the cutting line.
Since now we are not interested in counting the number of cut pomerons,
the sum over different bound state configuration gives just a global factor
$C^n$.
Performing the summation over the position of the cut $j$ we get
\bea
\sum_{j=1}^{2n-1} j (2n-j) S^n_j &=&
(-1)^n \sum_{1}^{2n-1} \frac{(-1)^j}{(j-1)!(2n-j-1)!} \nonumber \\
&=& (-1)^{n-1} \sum_{0}^{2n-2} \frac{(-1)^j}{j!(2n-j-2)!} \\
&=& \Bigg\{ \begin{array}{ll}
1      &   \textrm{if $n=1$} \\
0      &   \textrm{if $n>1$}
\end{array}\;, \nonumber
\eea
and we have the result
\beq
\tA^n_{1\textrm{-jet}}(s,t) = 0~~~~~~~   \textrm{if $n>1$}.
\eeq

\subsection*{$2$-Jet inclusive}
The situation is similar to the previous one: the combinatoric
factor $j(2n-j)$ is replaced by $j^2(2n-j)^2$. The separation in the
three cases analyzed in section \ref{2jets} can formally be 
obtained by writing one of the factors $j$ as $(j-1)+1$ and one of the factors
$2n-j$ as $(2n-j-1)+1$; the combinatorial factor $j^2 (2n-j)^2$ is therefore
written as a sum of four terms:
\beq
j^2(2n-j)^2=j(j-1)(2n-j)(2n-j-1)+j(2n-j)(2n-j-1)+j(j-1)(2n-j)+j(2n-j)\,.
\label{2jetdecomp}
\eeq
Let us also classify these terms counting the numbers ($n_l$,$n_r$)
of reggeized gluons on the (left, right)
side of the cut which emit at least one gluon (from the jet vertex).
If $n=1$ clearly only the last term in eq. \eqref{2jetdecomp} is
present (since $j=1$) and this correspond to the case $(n_l,n_r)=(1,1)$.
For $n\ge 2$ all the terms are not trivial:
\begin{itemize}
\item $j(j-1)(2n-j)(2n-j-1)$, corresponding to the jets being connected to 
different reggeized gluons on both sides, $(n_l,n_r)=(2,2)$;
\item $j(2n-j)(2n-j-1)$, corresponding to the jets being connected to 
different reggeized gluons on the right hand side but to the same gluon 
on the left hand side, $(n_l,n_r)=(1,2)$;
\item $j(j-1)(2n-j)$, corresponding to the jets being connected to different
reggeized gluons on the left hand side but to the same gluon on the right 
hand side, $(n_l,n_r)=(2,1)$;
\item $j(2n-j)$, corresponding to the jets being connected to the same
reggeized gluons on both sides, $(n_l,n_r)=(1,1)$.
\end{itemize}
It is easy to verify that after multiplication with the usual symmetry factor 
and summation over $j$, the last three term vanishes independently
and the first one vanishes unless $n=2$, so that the only surving
contributions appear when $n=(n_l+n_r)/2$.
 
\subsection*{$m$-jet inclusive}
We are now ready to generalize this result to the emission
of $m$ jets. The contribution of the $2n$-reggeized gluon diagram is
\beq
\tA^n_{m\textrm{-jet}}(s,t) =
2 i \sum_{\textrm{B.S.}}\sum_{j=1}^{2n-1} S^n_j j^m (2n-j)^m~
\int d\Omega_{2n}~s^{1+\tilde{\beta}}~ \cN_{2n}~ \cN_{2n}~
\tilde{\gamma}_{\{\bk_l\}}.
\label{discspQCDmjet}
\eeq
With the manipulations described in the previous subsection, we can
reduce this expression to a sum of terms, each of which corresponds to 
a diagram in which a certain number $n_r$ of reggeized gluons on the right
side of the cut emit at least one gluon, and $n_l$ reggeized gluons
on the left hand side emit one or more gluons. 
Each term is of the form

$$(-1)^n \sum_{j=n_l}^{2n-n_r} \frac{(-1)^j}{j!(2n-j)!}
j(j-1)\ldots(j-n_l+1) \cdot
(2n-j)(2n-j-1)\ldots(2n-j-n_r+1)$$
\bea
=& (-1)^n \sum_{n_l}^{2n-n_r} \frac{(-1)^j}{(j-n_l)!(2n-j-n_r)!}\nonumber \\
=& (-1)^n \sum_{0}^{2n-n_r-n_l} \frac{(-1)^{j+n_l}}{j!(2n-j-n_r-n_l)!} \\
=& \frac{(-1)^{n_l+n_r}}{(2n-n_l-n_r)!} \sum_{0}^{2n-n_r-n_l}
\binom{2n-n_l-n_r}{j}(-1)^j \nonumber \\
=& \Bigg\{ \begin{array}{ll}
(-1)^{\frac{n_l-n_r}{2}}     & \textrm{if $n=\frac{n_l+n_r}{2}$} \\
0                            & \textrm{if $n>\frac{n_l+n_r}{2}$}
\end{array}\;. \nonumber
\eea
Since $n_{l,r} \le m$, the inequality $(n_l+n_r)/2 \le m$ holds,
and all diagrams with $n>m$ vanish. Among the others, the non-vanishing ones
are those which satisfy the condition $n=(n_l+n_r)/2$. Briefly the
condition is
\beq
2n=n_l+n_r \, , \quad m>n
\label{jetcondition}
\eeq
and it fixes also the position of the cut to $j=n_l$.

The vanishing of \eqref{discspQCDmjet} for $n>m$ can also be seen in
another way:
\bea
\sum_{j=1}^{2n-1} S^n_j j^m (2n-j)^m &=&
\sum_{1}^{2n-1} \frac{(-1)^{n-j}}{j!(2n-j)!} j^m (2n-j)^m \nonumber \\
&=& \frac{1}{(2n)!} \sum_{0}^{2n}
\binom{2n}{j} (-1)^{n-j} j^m (2n-j)^m \nonumber \\
&=& \frac{(-1)^n}{(2n)!}
\frac{\partial^{2m}}{\partial \alpha^m \partial \beta^m} e^{2n\beta}
\underbrace{\sum_{0}^{2n} \binom{2n}{j} (-1)^{2n-j} e^{(\alpha - \beta) j}}_{
=(e^{\alpha-\beta}-1)^{2n}}|_{\alpha,\beta=0} \label{eq:espcomp} \nonumber \\
&=& \frac{(-1)^n}{(2n)!}
\frac{\partial^{2m}}{\partial \alpha^m \partial \beta^m}
(e^{\alpha}-e^{\beta})^{2n}|_{\alpha,\beta=0} \\
&=& \frac{(-1)^n}{(2n)!} \Big( x \frac{\partial}{\partial x} \Big)^m
\Big( y \frac{\partial}{\partial y} \Big)^m (x-y)^{2n}|_{x,y=1} \nonumber
\eea
The operator $(x\partial_x)^m$ can be written as
\beq
\Big( x \frac{\partial}{\partial x} \Big)^m =
\sum_{k=1}^{m} a_k^m x^k \frac{\partial^k}{\partial x^k}, 
\label{eq:opdec}
\eeq
where the coefficient $a_k^m$ are positive integer numbers whose explicit
expression is not needed. Using
\eqref{eq:opdec} in \eqref{eq:espcomp} we obtain
\bea
\frac{(-1)^n}{(2n)!} \sum_{k,k'=1}^{m} a_k^m a_{k'}^m x^k y^{k'}
\frac{\partial^k}{\partial x^k} \frac{\partial^{k'}}{\partial y^{k'}}
(x-y)^{2n}|_{x,y=1} = \nonumber \\
(-1)^n \sum_{k,k'=1}^{m} \delta_{k+k',2n} a_k^m a_{k'}^m (-1)^{k'}
\eea
which vanishes if $n>m$.

%%%%%%%%%%%%%%%%%%%%%%%%%%%%%%%%%%%%%%%%%%%%%%%%%%%%%%%%%%%%%%%%%%

\end{document}